\documentclass[11pt,preprint]{aastex}

\newcommand\be{\begin{equation}}
\newcommand\en{\end{equation}}

\shorttitle{$\eta$ Cha}
\shortauthors{Sicilia-Aguilar et al.}

\begin{document}

\title{The Long-Lived Disks in the $\eta$ Chamaeleontis Cluster}

\author{Aurora Sicilia-Aguilar\altaffilmark{1}, Jeroen Bouwman\altaffilmark{1}, Attila Juh\'{a}sz\altaffilmark{1}, Thomas Henning\altaffilmark{1},}
\author{Veronica Roccatagliata\altaffilmark{1}, Warrick A. Lawson\altaffilmark{2}, Bram Acke\altaffilmark{3,4}, }
\author{Eric D. Feigelson\altaffilmark{5}, A.G.G.M. Tielens\altaffilmark{6}, Leen Decin\altaffilmark{3}, Gwendolyn Meeus\altaffilmark{7}}
\altaffiltext{1}{Max-Planck-Institut f\"{u}r Astronomie, K\"{o}nigstuhl 17, 69117 Heidelberg, Germany}
\altaffiltext{2}{School of Physical, Environmental, and Mathematical Sciences, University of New South Wales, 
Australian Defense Force Academy, Canberra ACT 2600, Australia}
\altaffiltext{3}{Instituut voor Sterrenkunde, KU Leuven, Celestijnenlaan 200D, 3001 Leuven, Belgium}
\altaffiltext{4}{Postdoctoral Fellow of the Fund for Scientific Research, Flanders.}
\altaffiltext{5}{Department of Astronomy and Astrophysics, Pennsylvania State University, University Park, PA 16802}
\altaffiltext{6}{Kapteyn Astronomical Institute, PO Box 800, 9700 AU Groningen, The Netherlands}
\altaffiltext{7}{Astrophysical Institute Potsdam (AIP), An der Sternwarte 16, 14482 Potsdam, Germany}

\email{sicilia@mpia.de}

\begin{abstract}

We present IRS spectra and revised MIPS photometry for the 18 members of the $\eta$ 
Chamaeleontis cluster. Aged 8 Myr, the $\eta$ Cha cluster is one 
of the few nearby regions within the 5-10 Myr age range, during which the 
disk fraction decreases dramatically and giant planet formation must come to an end. 
For the 15 low-mass members, we measure a disk fraction
$\sim$50\%, high for their 8 Myr age, and 4 of the 8 disks lack near-IR 
excesses, consistent with the empirical definition of ``transition'' disks. Most of
the disks are comparable to geometrically flat disks.  
The comparison with regions of different ages suggests 
that at least some of the ``transition'' disks may represent the normal 
type of disk around low-mass stars. Therefore, their flattened structure and inner 
holes may be related to other factors (initial masses of the disk 
and the star, environment, binarity), rather than to pure time evolution.
We analyze the silicate dust in the disk atmosphere, finding moderate 
crystalline fractions ($\sim$10-30\%) and typical grain sizes $\sim$1-3~$\mu$m, 
without any characteristic trend in the composition.
These results are common to other regions of different ages, suggesting 
that the initial grain processing occurs very early in the disk lifetime ($<$1 Myr).
Large grain sizes in the disk atmosphere cannot be used as a proxy for age, 
but are likely related to higher disk turbulence. The dust mineralogy
varies between the 8-12~$\mu$m and the 20-30~$\mu$m features, suggesting high 
temperature dust processing and little radial mixing.  
Finally, the analysis of IR and optical data on the B9 star 
$\eta$ Cha reveals that it is probably surrounded by a young debris disk with a 
large inner hole, instead of being a classical Be star.

\end{abstract}

\keywords{accretion disks --- planetary systems: protoplanetary disks --- 
stars: pre-main sequence }

\section{Introduction \label{intro}}

The $\eta$ Chamaeleontis cluster is the nearest cluster discovered in the
20th century, and the first open cluster discovered by X ray observations 
(Mamajek et al. 1999). Located at 97 pc distance, and associated to the 
B9 star $\eta$ Cha, its age has been estimated to
be 5-9 Myr (Mamajek et al. 1999; Lawson et al. 2001; Luhman \& Steeghs 2004),
falling within the most interesting and less studied epoch,
when most disks disperse and giant planet formation has to conclude.
The $\eta$ Cha cluster is moreover a remarkable star-forming region. 
Its location out of the galactic plane and far from molecular clouds
raised questions about its origin, with the clusters space motions suggesting 
that it is probably related to the Sco-Cen star-forming clouds 
(Mamajek et al. 2000; De Zeeuw et al. 1999). The initial mass 
function (IMF) of the $\eta$ Cha cluster is unusual. Assuming an
IMF consistent with rich young clusters and field stars,
Lyo et al. (2004a) predicted 20-30 cluster members with
masses under 0.15 M$_\odot$, but deep photometric surveys
of large areas of the cluster that should have detected
brown dwarfs (BD) and planetary-mass objects revealed no very low-mass
members (Luhman 2004a; Lyo et al. 2006). This suggested either
an abnormal IMF or dynamical ejection of the very low-mass objects 
during the early phases of formation of the cluster (Lawson et al. 2001; 
Moraux et al. 2007). 

The $\eta$ Cha cluster has been one of the most
interesting targets for {\it Spitzer\,}, given its close distance,
which enables its members to be well characterized in
optical properties, spectral type, and binarity.
Although well-accepted evidence from other star 
forming regions suggested that most IR excesses from protoplanetary disks 
have disappeared by $\sim$6~My (Haisch et al. 2001), the $\eta$ Cha 
cluster appeared to have more accreting objects than expected from
ground-based JHKL photometry (Haisch et al. 2005). The presence of
a relatively large number of disks (disk fraction $\sim$50\%)
was later confirmed with Spitzer IRAC photometry by Megeath et al. (2005). 
The number of ``transition" objects (TO) or objects with
inner opacity holes (no or very small near-IR excess emission, having 
therefore evacuated or optically thin disks at these wavelengths) was 
larger than observed in Taurus or in massive star-forming 
regions (Megeath et al. 2005), explaining the initial
disagreement between near-IR excesses and accretion. 
Bouwman et al. (2006) used the IRS spectra of the low-mass
members to determine the disk fraction for single and
multiple stars, revealing that the lifetimes of disks
in close-in binary systems are significantly shorter than 
the lifetimes of disks around 
single stars ($\sim$5 versus $\sim$9~Myr, respectively). The
high disk fraction in the cluster is linked to the single stars,
as all but one of the binaries (the widest one, RECX-9) lack disks.
They also found that the angular momentum of the star is related
to the presence of disks and companions: the stars with disks
are slow rotators, and the fastest rotators are in binary systems. This 
is consistent with the disk locking hypothesis, as binaries would 
have had more time to spin up after losing their disks (Bouwman et al. 2006).

Here we study the disk properties of the cluster members, silicate dust
processing and evolutionary stage versus age, comparing to
similar low-mass star-forming regions, especially the
young Coronet cluster, also known as CrA ($\sim$1~Myr; 
Sicilia-Aguilar et al. 2008, from now on SA08, and references therein). The Coronet 
cluster shares with the $\eta$ Cha cluster a location, out of the 
galactic plane (De Zeeuw et al. 1999) and a population composed mainly of 
low-mass and very low-mass stars and BD, with only two intermediate-mass
members (spectral types BA).
Having very similar environment and population characteristics, 
the young CrA and the intermediate-aged $\eta$ Cha cluster seem 
to be ``twins'' at different stages of evolution. They are an ideal 
pair to study the effect of age on protoplanetary disks, separating 
the pure time evolution from the effects of the
environment, initial conditions, and stellar mass. In Section \ref{data}
we briefly introduce the IRS and MIPS observations. In Section \ref{analysis}
we study the disk structure and dust composition of individuals in
the global cluster context, examining the case of the B9 star
$\eta$ Cha itself. Finally, our results are summarized in
Section \ref{conclu}.

\section{Observations and Data Reduction \label{data}}

We obtained $7.5-35$~$\mu$m low-resolution ($R = 60-120$) of the $\eta$ Cha cluster members 
(see Table \ref{star-table})
with the Infrared Spectrograph (IRS, Houck et al. 2004) on-board the Spitzer Space Telescope.
A high accuracy optical peak-up (PCRS) was executed prior to the spectroscopic observations 
to position the target within the slit. Integration times were set such that the stellar 
photospheres could be detected with S/N$\sim$ 5. All targets have been observed with a 
minimum of three observing cycles for redundancy. 

Our spectra are based on
the {\tt droopres} products processed through the S15.3.0 version of the \emph{Spitzer} data
pipeline. Pixels flagged by the data pipeline as 
``bad'' were replaced with a value interpolated from an 8 pixel perimeter surrounding the errant 
pixel. Partially based on the {\tt SMART} software package (Higdon et al. 2004), our data were
further processed using the spectral extraction tools developed for the ``Formation and Evolution 
of Planetary Systems" (FEPS) {\it Spitzer\,} science legacy team (see also Bouwman et al. 2008).
The spectra were extracted using a 6.0 pixel and 5.0 pixel fixed-width aperture in the spatial 
dimension for the observations with the first order of the short- ($7.5-14$~$\mu$m) and the 
long-wavelength ($14-35$~$\mu$m) modules, respectively. The background was subtracted using 
associated pairs of imaged spectra from the two nodded positions along the slit, also eliminating 
stray light contamination and anomalous dark currents. 
The low-level fringing at wavelengths  $>20$~$\mu$m was removed using the
{\tt irsfinge} package (Lahuis \& Boogert 2003). To remove any effect of pointing offsets, we 
matched orders based on the point spread function of the IRS instrument, correcting for possible 
flux losses (see Swain et al. 2008 for further details). 

The spectra were calibrated using a 
spectral response function derived from multiple IRS spectra of the calibration star 
$\eta^1$~Doradus and a {\sc marcs} stellar model provided by the {\it Spitzer\,} Science Center. The 
spectra of the calibration target were extracted in an identical way as our science targets.
The relative errors between spectral points within one order are dominated by the noise on each 
individual point and not by the calibration. We estimate a relative flux calibration across an 
order of $\approx 1$~\% and an absolute calibration error between orders/modules of 
$\approx 3$~\%, which is mainly due to uncertainties in the scaling of the {\sc marcs} model.

The $\eta$ Cha cluster was mapped using the Multiband Imaging Photometer (MIPS) on
{\it Spitzer\,}. The observations were carried out on 2005 April 08 as part of 
the program number 100, using a medium scan map mode with half-array offsets.
The mapped area of around 0.5$^o$$\times$0.5$^o$ covered most of the known
$\eta$ Cha members. The MIPS photometry for the $\eta$ Cha members was
presented by Gautier et al. (2008), but we found a systematic disagreement
between their fluxes and our IRS observations that could not be explained
with the typical uncertainties in these high signal-to-noise data. Therefore,
we constructed the mosaics from the BCD data and repeated the photometry as
standard aperture photometry. We analyzed the most recent calibrated data 
(BCD version S16.1.0).
The mosaics at 24 and 70~$\mu$m were created using the MOPEX
software package (version 16.2.1). Taking into account the typical overlap
per channel, the total exposure times were of 107s and
54s at 24 and 70~$\mu$m, respectively. We used apertures
of 13'' and 30'' for the 24 and 70~$\mu$m photometry, respectively. 
The sky was measured in an annulus between  20''-32'' for 24~$\mu$m
and 40''-60'' for 70~$\mu$m, and we applied aperture corrections
of 1.167 and 1.295 (for more details, see the MIPS Data 
Handbook\footnote{http://ssc.spitzer.caltech.edu/mips/dh}). 
The measured MIPS magnitudes are listed in Table
\ref{IR-table}. The uncertainties represent only the 1$\sigma$
statistical error without including further potential sources of errors that
may sum up to $\sim$10\% of the flux, depending on the location 
within the chip, flat fielding, and calibration
(see {\it Spitzer\,} MIPS manual). The agreement between the IRS fluxes and our
newly calculated MIPS magnitudes is very good, with typical errors within
few percent.

\begin{figure}
\plotone{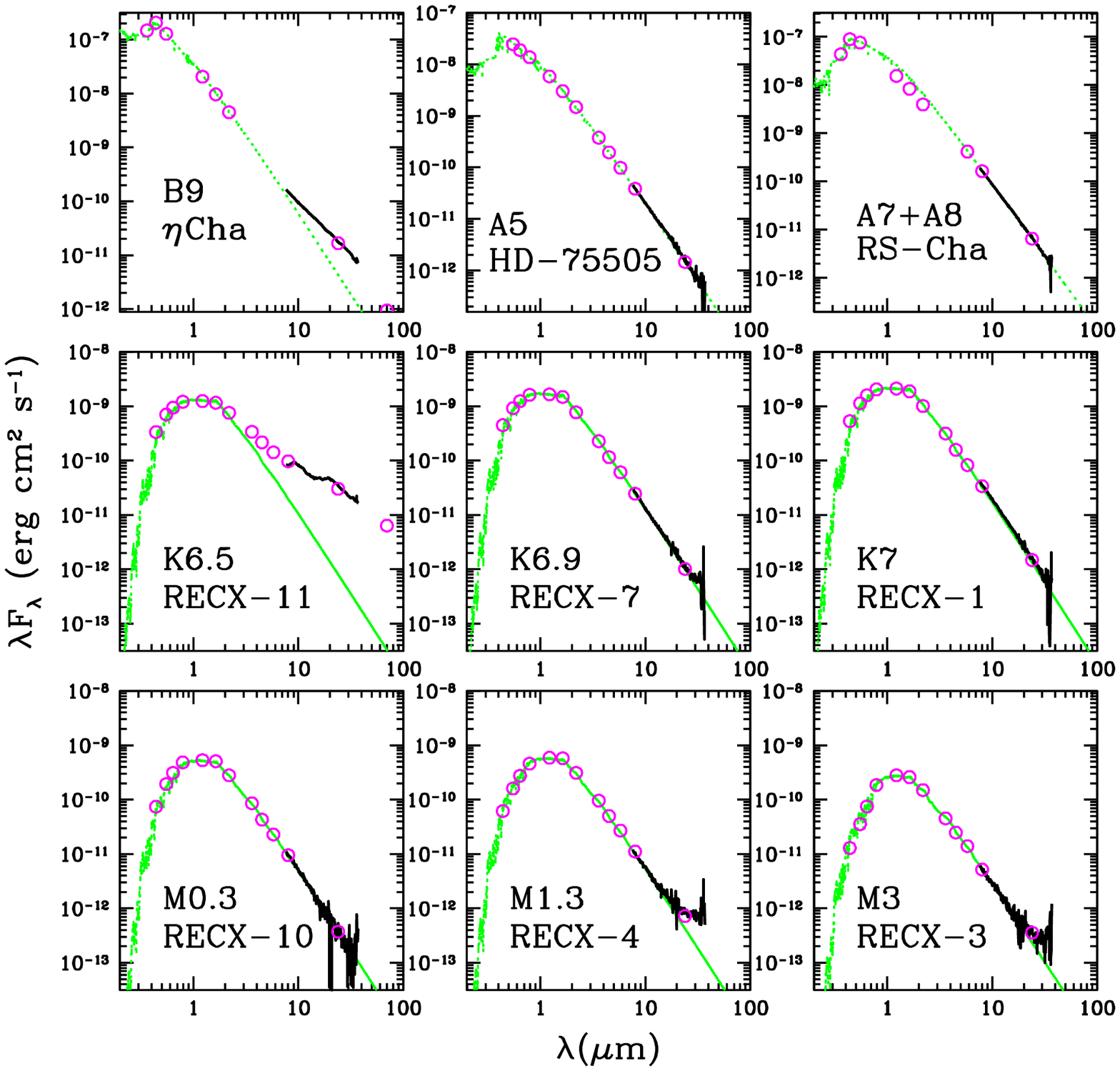}
\caption{SEDs of the $\eta$ Cha members, including optical and IR data
(open circles), IRS spectra (thick black line) and stellar models (dotted green line,
including {\sc marcs} models for the low-mass stars).
(see Tables \ref{optlit-table} and \ref{IR-table}). RS Cha is an eclipsing binary; 
the 2MASS data were obtained during an eclipse and accordingly fall under the photospheric 
model which fits the combined SED of both stars\label{seds1-fig}}
\end{figure}

\begin{figure}
\plotone{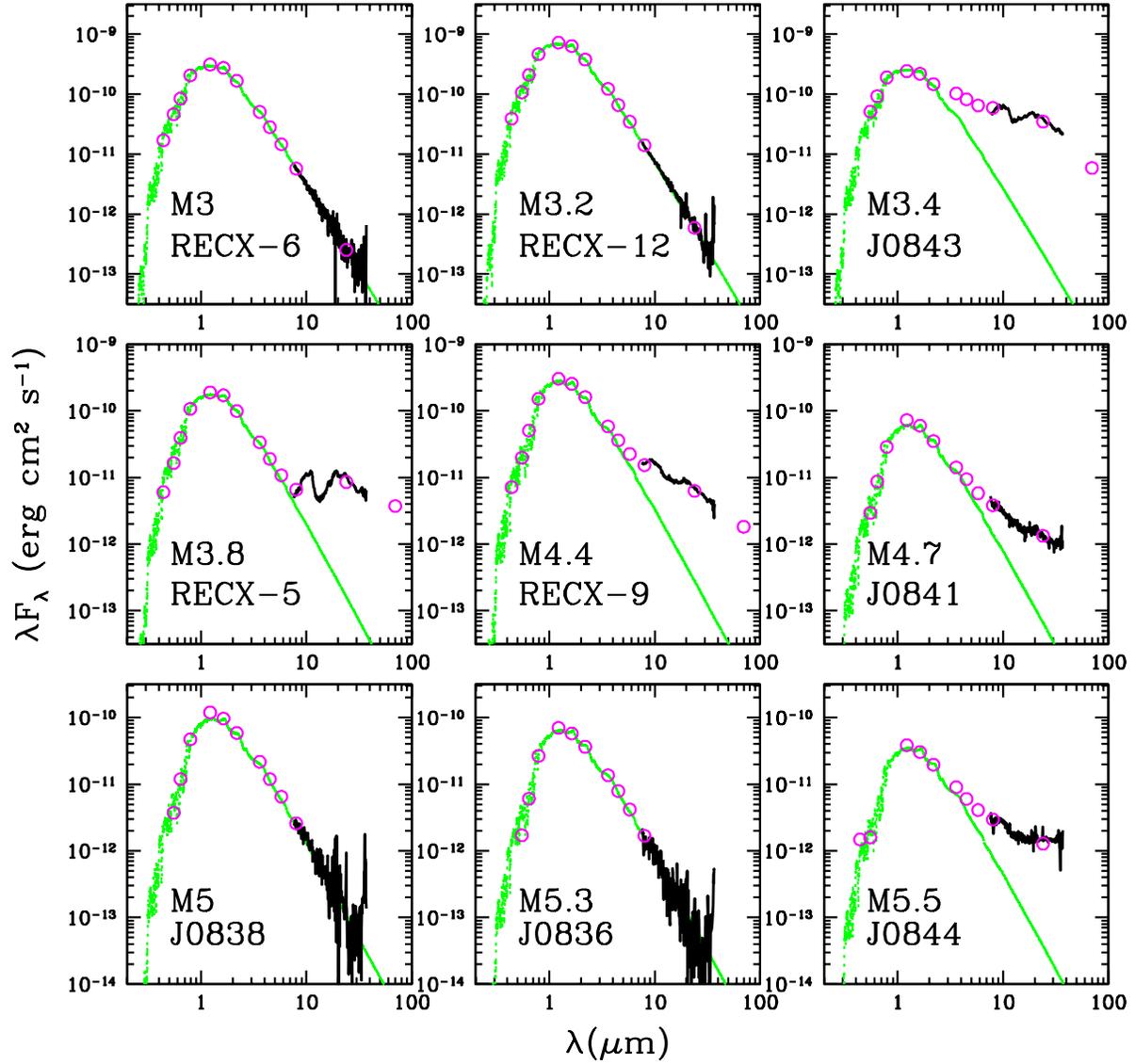}
\caption{SEDs of the $\eta$ Cha members, including optical and IR data
(open circles), IRS spectra (thick black line) and stellar models (dotted green line,
including {\sc marcs} models for the low-mass stars).
(see Tables \ref{optlit-table} and \ref{IR-table}).
\label{seds2-fig}}
\end{figure}

\begin{figure}
\plottwo{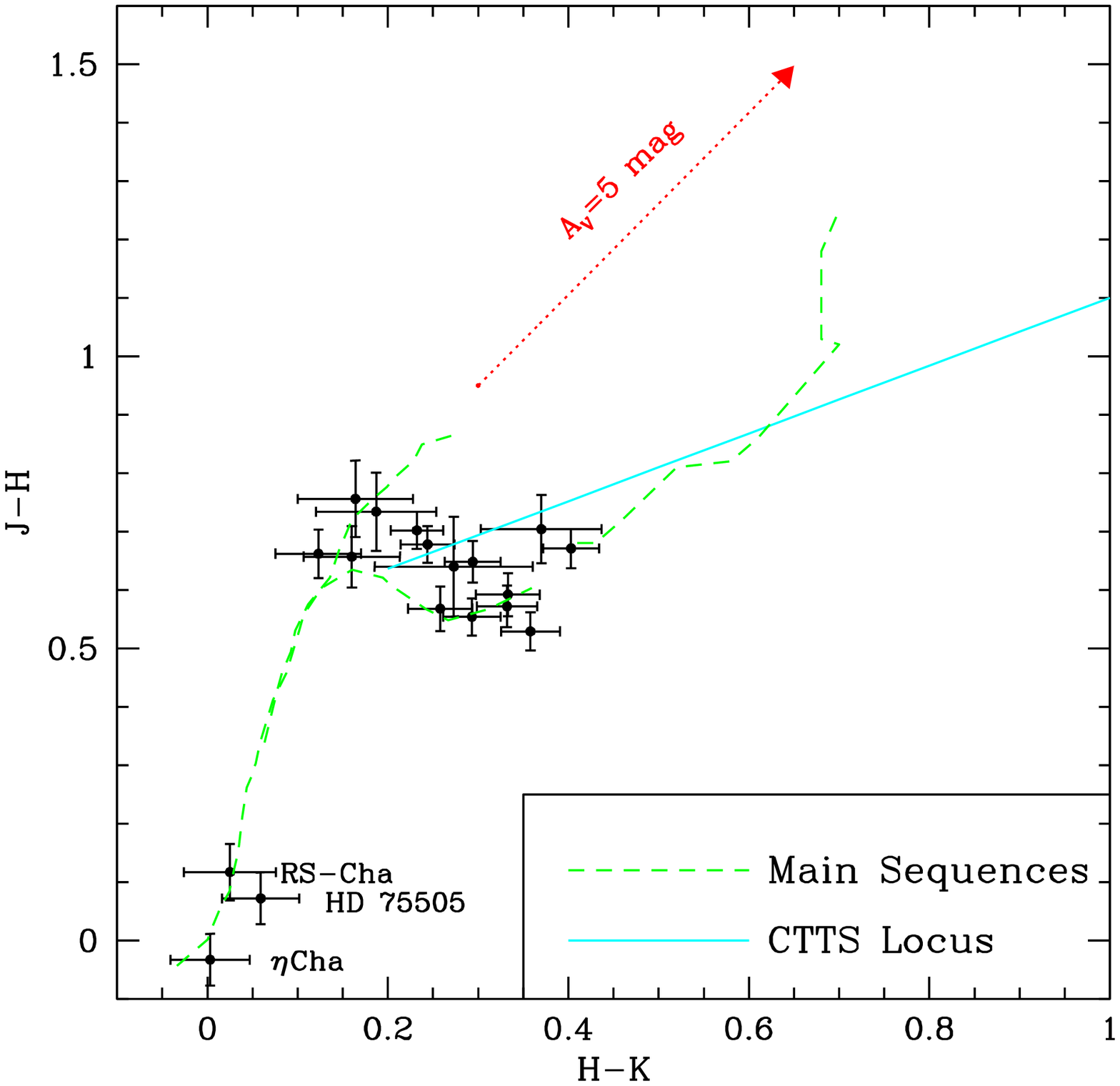}{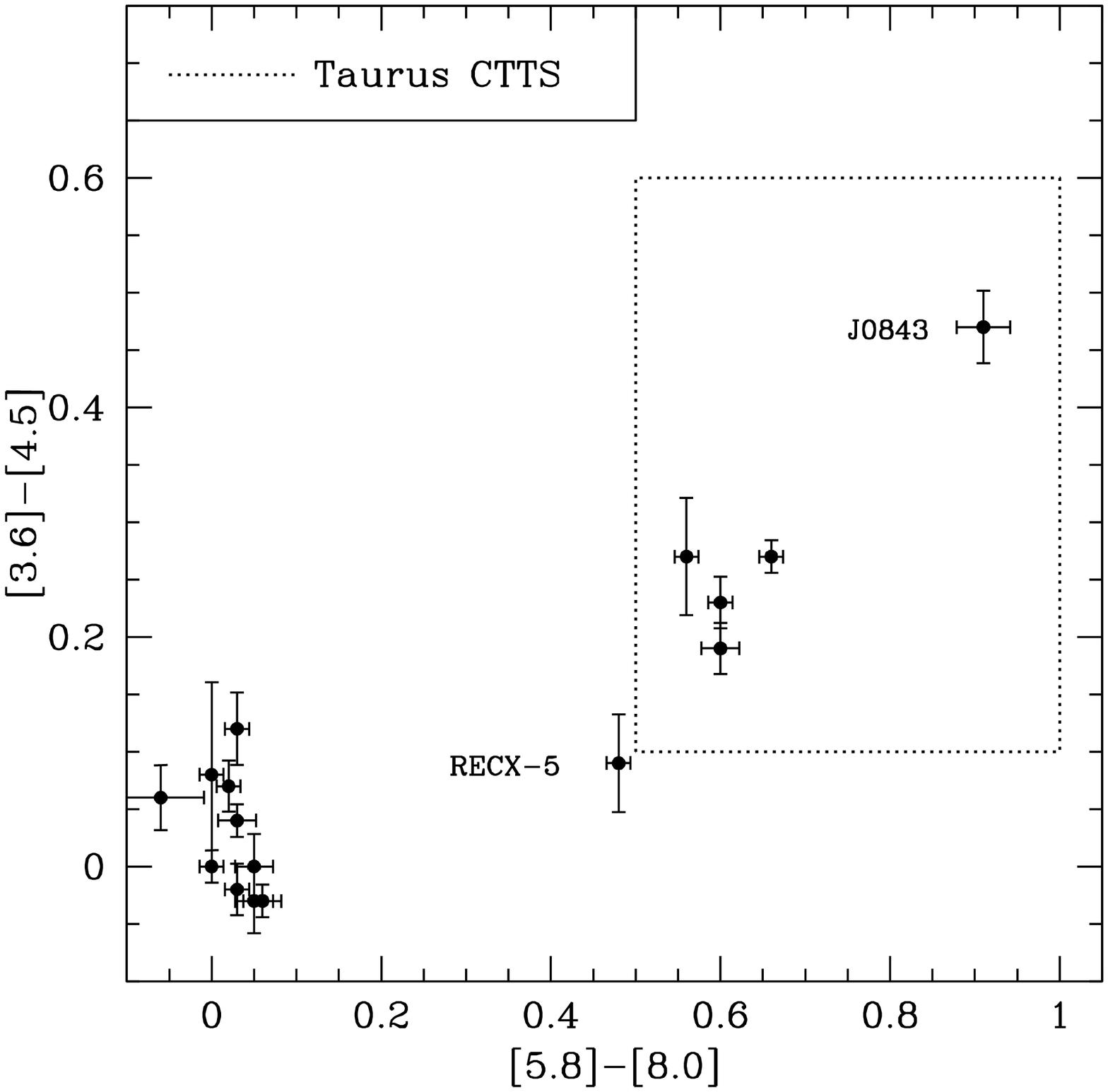}
\caption{2MASS and IRAC color-color diagrams for the $\eta$ Cha cluster. 
The JHK diagram shows the main sequences for dwarfs, giants (Bessell \& Brett 1989), 
and brown dwarfs (Kirkpatrick et al. 1996), and the locus of the classical T Tauri stars 
(CTTS locus; Meyer et al. 1997). RECX-5 shows
typical colors of a TO with an excess starting after $\sim$5.8~$\mu$m, and
J0843 presents a strong excess typical of a very flared disk. 
The four other object with colors consistent with Taurus CTTS are RECX-9,
RECX-11, J0841, and J0844, but since their disks are flatter/transitional, the objects
are less red than the typical Taurus CTTS. \label{irac-fig}}
\end{figure}

In order to complete the spectral energy distributions (SEDs; 
Figures \ref{seds1-fig} and \ref{seds2-fig}) and to compare the IR excess emission with
the presence of gas accretion, we also include the data available for
the cluster members in the literature (see Tables \ref{star-table}, \ref{optlit-table}, and 
\ref{IR-table}). Optical magnitudes and spectral types are taken from Johnson et al. (1966),
Mermilliod (1991), Mamajek et al. (1999), Lawson et al. (2001, 2002, 2004), 
and Lyo et al. (2003, 2004a, 2004b). Near-IR magnitudes from 2MASS (Cutri et al. 2003) 
are also available, as well as IRAC magnitudes (Megeath et al. 2005).
Figure \ref{irac-fig} shows the JHK and IRAC color-color diagrams for all the
$\eta$ Cha cluster members.

Finally, for the purpose of modeling the underlying stellar atmosphere, a small grid of 
theoretical spectra was generated using the {\sc marcs} model atmosphere code (version 1998, 
Plez et al., 1992, and references therein). Detailed photospheric models, including IR opacities, 
are particularly important to study the excesses over the photospheric emission, which may be 
very small for the typical flattened disks around low-mass stars.
The {\sc marcs}-code is aimed at modeling the 
atmospheres of cool stars, allowing for both plane-parallel and spherical geometries. Basic 
assumptions are local thermodynamic equilibrium (LTE), hydrostatic equilibrium, and conservation 
of energy for radiative and convective flux. The SEDs 
for the {\sc marcs} model atmospheres were calculated with the {\sc turbospectrum} code (Plez et al. 1992) 
using the same physical input parameters of atomic and molecular equilibrium constants, solar 
abundances, and continuous opacity source, as described in Decin et al. (1998, 2000). 
For the line opacities, an infrared database including CO, SiO, 
H$_2$O, OH, NH, HF, HCl, CH, and NO was prepared (references and discussion
can be found in Decin 2000). The grid of stellar spectra was generated for 
effective temperatures between 3200 and 4100\,K (in steps of 300\,K) and gravity values 
($\log$ g [cm/s$^2$]) of 4.0, 4.5, 5.1, 5.6, and 6.10. A solar metallicity was assumed and the 
computations were performed for a plane-parallel geometry due to the high gravity values. The 
synthetic spectra were derived for a resolution of 0.1\,\AA\ and then degraded to the IRS
resolution (R$\sim$100). The {\sc marcs} models for the low-mass stars are displayed,
together with the data, in Figures \ref{seds1-fig} and \ref{seds2-fig}.

\section{Analysis and discussion\label{analysis}}

\subsection{Disks around low-mass stars: Are the ``transition'' disks really in transition? \label{to}}

Among the 15 low-mass members in the $\eta$ Cha cluster (spectral types K5.5-M5.5),
we find 8 objects with IR excesses. Two of them (RECX-11, J0843) display the typical SEDs 
of classical T Tauri stars (CTTS) or Class II objects with flared disks, consistent 
with the Taurus median SED (including quartiles). 
Two more (J0841, J0844) are Class II objects with 
optically thick, but geometrically thin (flattened) disks (SED slope
$\lambda$F$_\lambda \sim \lambda^{-4/3}$; Hartmann 1998)
and maybe even optically thin inner disks. Four objects (RECX-3, RECX-4, RECX-5, RECX-9)
have inner opacity holes (no excess at wavelengths shorter than 
6~$\mu$m), being empirically classified as ``transition'' objects (TO). The remaining 
7 members have no significant IR excess, 
being Class III objects (see Figures \ref{seds1-fig}, \ref{seds2-fig}, \ref{irac-fig}). 

Detecting inner holes in very low-mass stars with cold photospheres is more 
uncertain than in solar-type objects (spectral type K-early M), as the 
disk IR excesses tend to start at longer wavelengths
and depend strongly on the size and angle of view of the disk inner rim (as
it happens for solar-type stars with the K band excess). Therefore, we label
as TO those objects with no significant excess at $\lambda <$6~$\mu$m. 
This is an observational criterion, and it may not be sensitive to very small 
inner holes around the higher-mass stars. Some of the most flattened 
disks around late-M stars may have very small or negligible excesses around 6~$\mu$m 
despite having optically thick disks without holes, as recent models have shown 
(Ercolano et al. 2009). Nevertheless, the contrast between solar-type and
M-type stars is less dramatic if we assume more realistic temperatures
for both types of objects, e.g. $\sim$4400 K for 1 Myr-old solar-type 
stars (a young, $\sim$1 M$_\odot$ star has an spectral type $\sim$K5-K6)
and $\sim$3400 K for the M-type stars (e.g., Siess et al. 2000), instead 
of the values of 5000 K and 3000 K used by Ercolano et al. (2009).
Flattened disks have historically been 
labeled as TO as well, given their differences with the typical, flared disks 
(e.g. Lada et al. 2006; Mer\'{\i}n et al. 2008). Currie et al. (2009) defined
these objects with flattened disks as ``homologously depleted'' transition disks, 
which should be considered as part of the TO family, given that their reduced
IR excess involves substantial disk evolution with respect to flared
Class II objects. This classification would include objects like J0841 in
the TO family, and J0844 would be a border-line case. In addition, it is not 
known whether all disks 
evolve from Class II to Class III objects via the opening
of an inner hole, and evolution may occur over a larger radial distance.

Two of the TO have small excesses (RECX-4 and RECX-3) that could also
correspond to optically thin ``debris'' disks (Kenyon \& Bromley 2005).
There are so far no statistical studies of the luminosity and
evolution of debris disks around very low-mass stars. Some studies,
based on millimeter data, suggest that debris disks are as common for M-type 
stars as they are for intermediate-mass stars (e.g., Lestrade et al. 2006),
but they are based on a small number of objects, typically much older than
the $\eta$ Cha cluster members. Distinguishing debris disks from
TO for M-type objects with IR observations is not exent of problems. Unlike in
more massive stars, the processes resulting in dust removing (e.g., Poynting-Robertson
drag) require longer time to clean the disk ($\sim$1~Myr), so objects with small
excesses do not necessarily have second-generation dust (Currie et al. 2009). 
Therefore, within the limitations of small-number statistics, the estimated 
disk fraction is $\sim$50\%, and about half of the disks are TO, having probably 
cleaned or optically thin inner disks. 

This disk fraction is similar to the values observed in younger clusters like
the 4 Myr-old Tr 37 ($\sim$50\%; Sicilia-Aguilar et al. 2006a\footnote{Note that
this study is not complete at 24~$\mu$m, so objects with large holes similar to RECX-3 and
RECX-4 are not included in the disk fraction for Tr 37. This can be generalized
to many of the existing disk surveys, including those mentioned in Section \ref{evol}. 
Complete surveys at 24~$\mu$m would probably reveal larger disk fractions for these 
regions.}), and
higher than what we would expect for a $\sim$8 Myr-old
cluster (Haisch et al. 2001; Hern\'{a}ndez et al. 2007). Nevertheless, most
studies have concentrated on solar-type stars, and in any case, there is not 
much information about clusters older than 4 Myr. The fact that most of 
the low-mass stars in the $\eta$ Cha cluster have spectral types later than
M0 could explain differences in disk mass, disk lifetimes, disk fractions, and disk
structure/morphology if these parameters are related to the stellar mass. 
In addition, most of the clusters that have been
analyzed to trace the evolution of the disk fraction are massive 
star-forming regions, containing several high- and intermediate-mass stars and
hundreds of low-mass stars. The  $\eta$ Cha cluster is very different, as it contains 
only 18 known members, most of them low-mass stars. If the environment and/or 
initial conditions have an influence in the formation of the disks
and their subsequent evolution, we may expect differences between the
disks characteristics and disk fraction in high-mass versus low-mass 
clusters.

TO represent usually only a small fraction ($\sim$5-10\%) of the total number of disks
(Hartmann et al. 2005; Sicilia-Aguilar et al. 2006a; Hern\'{a}ndez et al. 2007).
Nevertheless, our study of the 
young ($\sim$ 1 Myr) Coronet cluster (SA08) revealed also 
a large fraction of TO ($\sim$50\% of the total number of disks applying our 
observational criterion and including ``homologously depleted'' TO). The 
combined results of the $\eta$ Cha and Coronet clusters suggest 
that disks with opacity inner holes may be as (or nearly as) frequent as CTTS-like,
optically thick disks for these late-type TTS at any age.
Recently, it has been suggested that our high TO fraction in the
Coronet cluster was inaccurate (Ercolano et al. 2009), and most of
the disks could be fitted by optically thick flattened SEDs without inner
holes, leaving a TO fraction not different from observed around solar-type
stars. This study provides valuable informtion that could be also applied to
our results in the $\eta$ Cha cluster. While both the Coronet cluster and
the $\eta$ Cha cluster are limited by low-number statistics given their
small populations, we believe that TO fractions are higher than for solar-type
stars, even if the SEDs of some objects may indeed be reproduced by models without holes.
In particular, we strongly disagree with the model presented for 
CrA-205 by Ercolano et al. (2009), as it ignores the near-IR data on the
basis of low S/N, and predicts a flux at $\sim$8~$\mu$m that is more than 5$\sigma$
higher than both our IRS and IRAC observations. Their models for several
objects (G-14, CrA-4111, and also CrA-205) require as well stellar
luminosities and extinctions that differ from the observed values by
more than the typical uncertainties, which suggests further model limitations. 
Therefore, we obtain a
TO fraction of at least $\sim$30\%$\pm$10\% for the Coronet cluster, 
higher if we include the homologously depleted ``transition'' objects 
and the objects with small excesses at $\lambda >$20~$\mu$m 
(G-30, G-94, G-95, and G-102). 

A high fraction of TO seems to be present as well
in the MBM12 cluster (Meeus et al. 2009). With a small population
(like $\eta$ Cha and the Coronet cluster), an age of 2 Myr, and a
very high disk frequency ($\sim$90\%), about 40\% of the disks in the MBM12 cluster
resemble TO. The high number of flattened/transitional
disks seems then ubiquitous for low-mass stars in small clusters of different
ages, suggesting that their lifetimes are comparable to the lifetimes of the 
CTTS-like, optically thick disks. Therefore, at least some of the TO may not be in
a short-lived stage between Class II and Class III objects, but may
spend a large part of their lives as such structures. 

The main reason to treat TO as something different from CTTS-like protoplanetary 
disks is their assumed short life. Since usually only 5-10\% of the disks around 
solar-type stars are TO, their lifetimes are thought to be 
shorter than the lifetimes of ``normal'' protoplanetary disks.
In such scenario, TO would be in a rapid, ``transitional''
stage between Class II and Class III objects, and the whole disk
would disperse within 10$^5$-10$^6$ yr after the innermost
disk has been cleared. This could be achieved either by photoevaporation (Alexander et al. 2006a,b)
and/or by  planet formation (Quillen et al. 2004). 
More recently, it became
evident that, although some TO harbor planets (e.g. TW Hya; Setiawan et al.
2008), others may have their inner part cleared since their formation
due to the presence of close-in binaries (e.g. CoKu Tau/4; Ireland \& Kraus 2008).
Although the structure of disks 
around M-type stars and even BD is sometimes considered as a smaller-scale version 
of solar-type TTS, disk masses around very low-mass objects
are not well constrained, and the properties and structure of the 
innermost disk could be different if the disk masses are lower than 
in disks around solar-type stars (Hartmann et al. 2006). 
Therefore, the ``transition'' morphology may be due in some cases to other causes,
instead of being an evolutionary 
process in an initially ``normal'' disk. In this case,
the characterization of the properties of TO is needed to understand the disk physics. 

The accretion rates can be used to test the planet formation and 
photoevaporation scenarios for TO. Considering the disk masses and accretion 
rates of the TO in Taurus, Najita et al. (2007)
concluded that most of them are consistent with inner disk clearing by 
the formation of Jupiter-sized planets, being roughly one order of magnitude
lower than the typical accretion rates for Class II objects in the region 
(Lubow \& D'Angelo 2006). The two accreting TO in $\eta$ Cha (RECX-5 and RECX-9) 
have very low accretion rates (see Table \ref{star-table}). Nevertheless, these
accretion rates are not significantly different from those of the accreting
Class II objects (J0843 and RECX-11), so in this case we cannot rule out nor
confirm the planet formation scenario. On the other hand, the accretion rate
of RECX-11 (\.{M}$\sim$4 10$^{-11}$M$_\odot$/yr) is well under the limit at which 
photoevaporation should become efficient to clean the innermost disk in short 
time (\.{M}$\sim$4 10$^{-10}$M$_\odot$/yr; Hollenbach et al. 2000). 
The fact that RECX-11 does not present an inner hole suggest 
that either the UV luminosities of young stars have been
overestimated, or that the efficiency of photoevaporation is somehow reduced
(for instance, because of gas shielding by the dust).

Regarding the presence of stellar companions to explain the TO, 
we do not have information about the Coronet cluster members, but several
$\eta$ Cha members are known to be multiple systems (Table \ref{star-table}). 
In $\eta$ Cha, 
Bouwman et al. (2006) found that the presence of disks and nearby companions are 
anticorrelated, as the lifetimes of disks around binary systems are lower
than for single stars. RECX-9 is probably the widest binary in the cluster, and the only
with a (circumprimary) disk. Though the projected separation 
is only 20 AU, no motion has been detected in the system (Bouwman et al. 2006). 
A previous low-resolution spectroscopic study found a spectroscopic 
binary around RECX-7 with a period of 2.6 days (Moraux et al. 2007; Lyo et al. 2003).
Nevertheless, the presence of  spectroscopic binaries (which are the ones that could 
be responsible for the holes with sizes of few tenths of AU to a few AU detected 
at {\it Spitzer\,} wavelengths) around the rest of the $\eta$
Cha cluster members has not been explored yet.

\subsection{Evolution of the disks SEDs for low-mass objects \label{evol}}

\begin{figure}
\plottwo{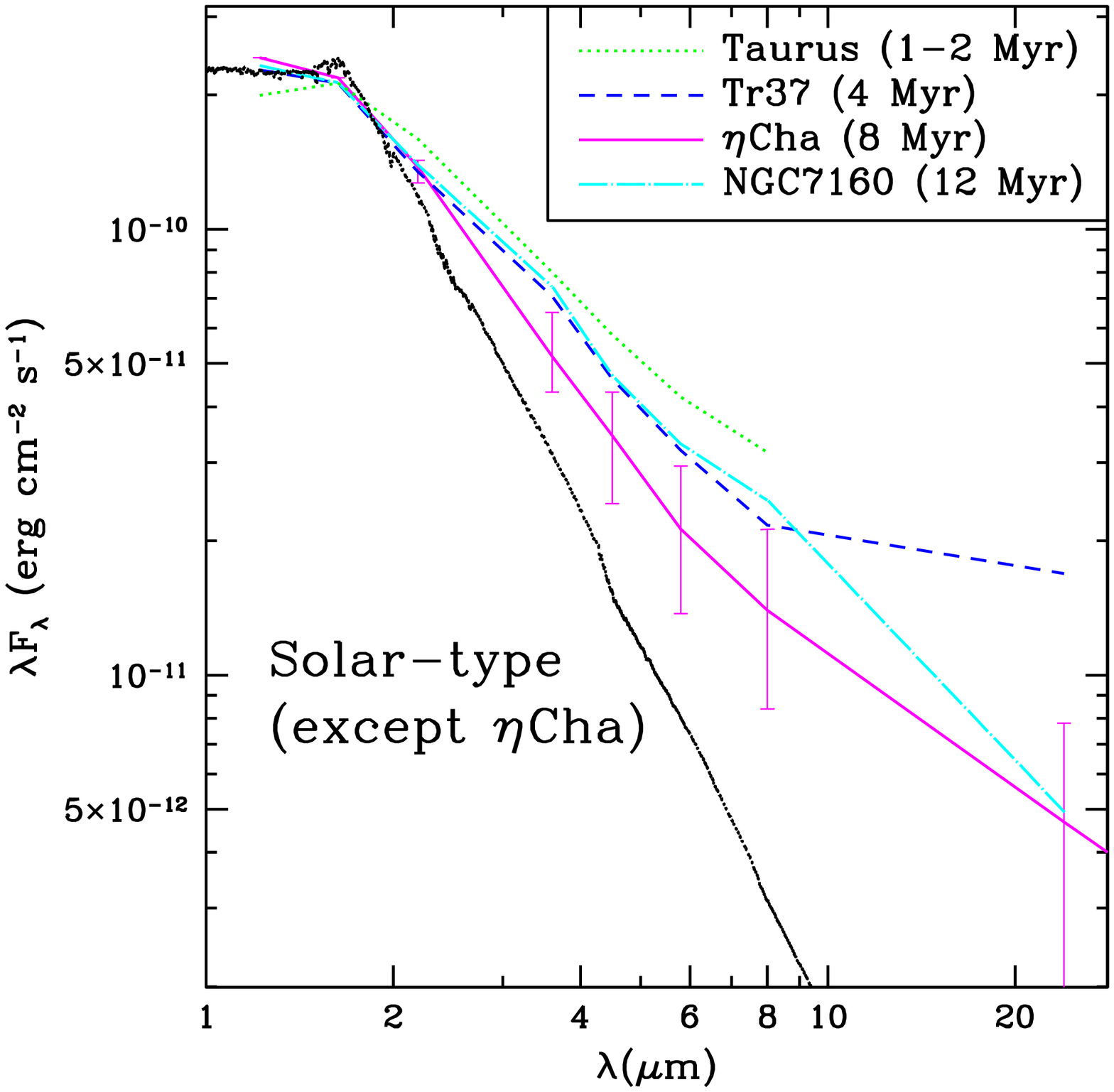}{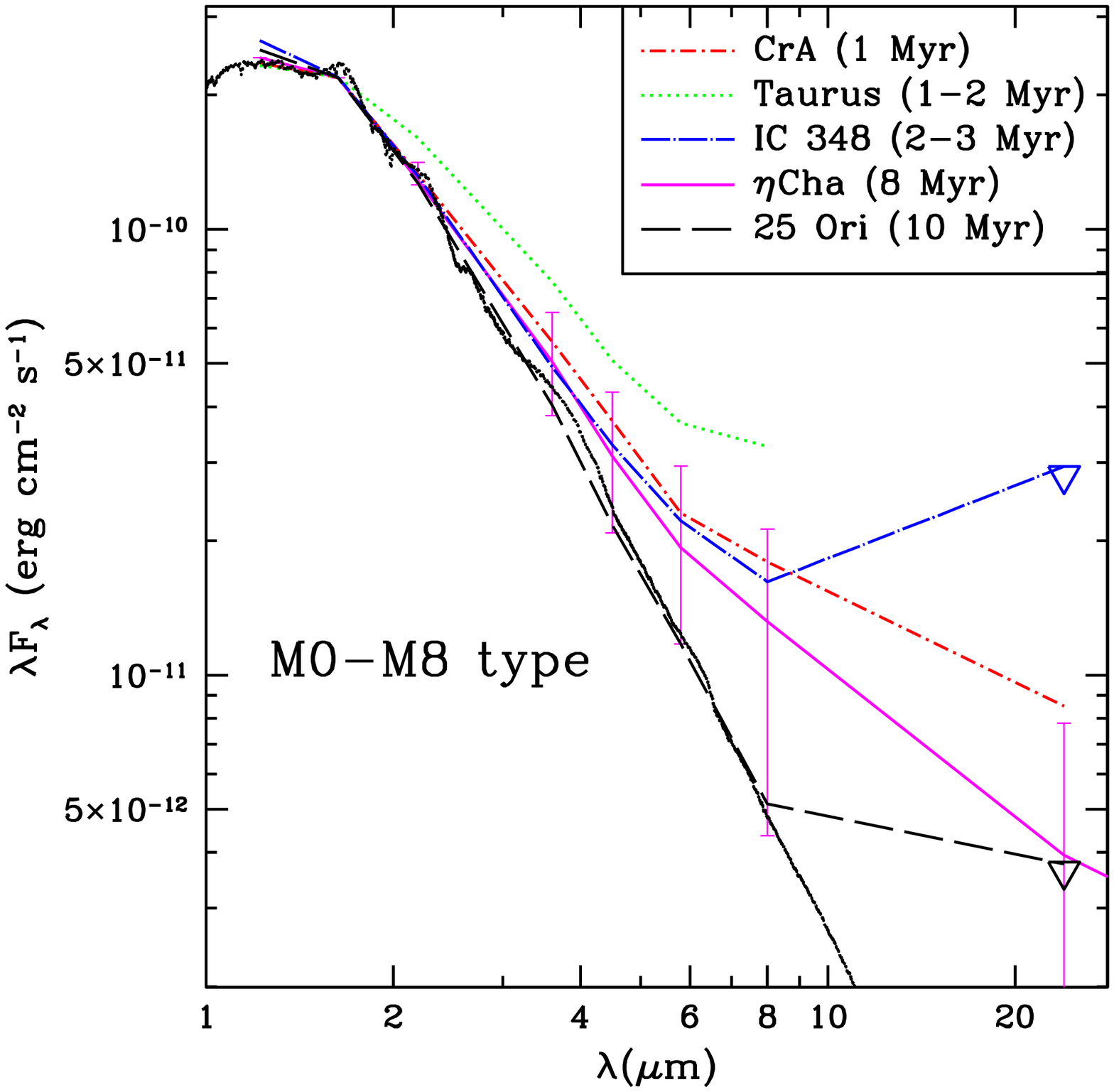}
\caption{Median SED for $\eta$ Cha low-mass (M-type) members, compared to the median SEDs of different
regions for solar-type (K-early M, left) and M-type objects (right) with different ages:
Tr 37 and NGC 7160 (4 and 12 Myr, respectively; Sicilia-Aguilar et al. 2006); 
CrA (1 Myr; SA08);
Taurus (1-2 Myr, Hartmann et al. 2005); IC 348 (2-3 Myr; Luhman et al. 2003; Lada et al. 2006);
25 Ori (10 Myr; Brice\~{n}o et al. 2007; Hern\'{a}ndez et al. 2007).
All the colors have been dereddened using the corresponding extinctions, and the fluxes have
been scaled to match the distance and luminosity of the $\eta$ Cha cluster. 
The median SEDs are compared to the photospheric emission of a K7 (left) and an M4 star (right, 
also from the {\sc marcs} models; dotted line).
All the datasets are complete at IRAC wavelengths, MIPS 24~$\mu$m is only complete for
the $\eta$ Cha cluster and the Coronet cluster (see text); therefore it is marked with
an inverted triangle as an upper limit in IC 348 and 25 Ori. The quartiles
are displayed for the $\eta$ Cha cluster, being similar for the other regions.
\label{mediansed-fig}}
\end{figure}

The differences in disk morphology of the $\eta$ Cha cluster members compared to
other TTS disks are reflected in the
typical spectral energy distributions (SEDs), which are very different from what is
observed for solar-type stars in other regions (Sicilia-Aguilar et al. 2006a;
Figure \ref{mediansed-fig} left). The median SED of the $\eta$ Cha disks around
low-mass members (Figure \ref{mediansed-fig} right) strongly differs from the typical
SED for Taurus disks, falling below the SED of
an optically thick, geometrically thin disk ($\lambda$F$_\lambda \sim \lambda^{-4/3}$;
Hartmann 1998). This is indicative of a high degree of dust evolution/disk
cleaning in the innermost disk. A similar median slope is also found in 
the young Coronet cluster (Figure \ref{mediansed-fig} right; SA08). 
The median spectral type for objects with disks in both the $\eta$ Cha and 
the Coronet clusters is M3-M4. 

In order to reveal the possible differences in disk evolution versus stellar mass 
and environment, we traced the median disk SED for M stars in several regions 
for which accurate spectral types, extinctions, and {\it Spitzer\,} data were 
available: the Coronet cluster (13 objects, median spectral type M4, $\sim$1 Myr; SA08), 
Taurus (23 objects, median spectral type M3, $\sim$1-2 Myr; Kenyon \& Hartmann 1995; 
Brice\~{n}o et al. 1997; Hartmann et al. 2005), IC 348 (128 objects, 
median spectral type M4,  $\sim$2-3 Myr; Luhman et al. 2003; Lada et al. 2006),  
and 25 Ori (4 objects, median
spectral type M3, $\sim$10 Myr; Brice\~{n}o et al. 2007; Hern\'{a}ndez et al. 2007). 
For each region, we considered all the objects with IR excesses and known
spectral types (which are the majority of the disk members), including TO.
The IRAC data on all these regions are complete (detecting diskless photospheres
in all the mass range). The MIPS 24~$\mu$m data are complete only for the $\eta$ Cha
and the Coronet clusters. Therefore, the median points at 24~$\mu$m for IC 348 and
25 Ori are upper limits, which may explain why the IC 348 24~$\mu$m value is much higher 
here than in the other regions. 

The median SEDs were computed after correcting for the individual extinction
of all objects in each region. For 
Taurus and 25 Ori, we used the average A$_V$=1 mag
and A$_V$=0.29 mag, respectively, for the objects with unknown extinction. 
Since the typical extinctions are low in both regions, individual variations are 
not significant at IR wavelengths.  
The median SEDs for low-mass objects (M0-M8)
are displayed in Figure \ref{mediansed-fig} (right). All the regions, except 
Taurus, show a median SED at $\lambda <$6~$\mu$m consistent with or lower than
an optically thick,  geometrically thin disk ($\lambda$ F$_\lambda \sim \lambda^{-4/3}$). 
For 25 Ori, 3 of the disks are TO with no excess at  $\lambda <$8~$\mu$m.
Solar-type objects in the same regions have different median disk SEDs 
(Lada et al. 2006), which is consistent with differences in disk structure and/or 
evolution depending on the spectral type. 

This suggests that flattened (geometrically thin) and ``transition'' 
disks are more common among the low-mass
objects than among solar-type stars in most regions, as previously suggested by
Megeath et al. (2005), McCabe et al. (2006), and our observations of early M-type
stars in the Cep OB2 region (Sicilia-Aguilar et al. 2007). 
One reason for this behavior could be that
a given IR wavelength traces a smaller and closer-in zone in a mid- or late-M
star than in a solar-type star, so any signs of inner disk evolution (grain growth,
flattening, inner holes) will drastically reduce the IR excess
(Kessler-Silacci et al. 2007). 
On the other hand, the fact that the median SEDs of $\eta$ Cha and the Coronet
cluster are not significantly different at $\lambda <$10~$\mu$m, and very flat
at all wavelengths, suggests
that the inner disk structure for the disks around low-mass objects 
differs from their solar-type counterparts at all ages. Differences in the
disk structure (maybe because of the lower disk masses and the
lack of dead zones; Hartmann et al. 2006) and/or initial
conditions for the low-mass objects (for instance, in the angular momentum of the 
cores that form very low-mass stars; Dullemond et al. 2006)
may be responsible for this effect. In addition, 
the environment (high-mass versus low-mass star forming regions) and the
initial conditions in the star-forming cloud could affect
both the formation and the evolution of the objects and their disks. This
might explain why regions like Taurus (containing a relatively numerous population
of low-mass stars in an undisturbed, quiescent environment, compared to OB
associations) deviate from the flat-disk general trend seen in other clusters.

\subsection{Dust mineralogy in the disk atmosphere: Grain processing and radial mixing \label{extract}}

The optically thin silicate emission seen at $\sim$10~$\mu$m and $\sim$20~$\mu$m
in the IRS spectra traces the dust in the optically thin, warm (T$\sim$150-450 K)
disk atmosphere (Calvet et al. 1992). Although this is only a very small part of the 
total dust contained in the disk, the presence of processed grains in the disk 
atmosphere reveals grain evolution (compared to ISM dust), grain transport in the disk, 
and the effect of turbulence. Among the $\eta$ Cha cluster members, only 5 show
silicate emission over the continuum at 10~$\mu$m, and from those,
only 4 display a clear silicate feature with good enough S/N 
to allow a detailed analysis (RECX-5, RECX-9,
RECX-11, and J0843). The silicate feature is weak in J0844, and
absent in J0841.  The spectrum of $\eta$ Cha does not show any
features. There are no evident gas lines in the $\sim$13-14~$\mu$m region
(Carr \& Najita 2008) at more than a 3$\sigma$ level. 

To study the grain properties in the disk atmospheres, 
we used the Two Layer Temperature Distribution (TLTD)
spectral decomposition routines (Juh\'{a}sz et al. 2009, hereafter J09). 
The model reproduces the silicate 
emission using the sum of a multicomponent continuum and the optically thin emission 
of a set of 4 different dust species (amorphous silicates
with olivine and pyroxene stoichiometry, forsterite, silica) 
with sizes 0.1, 1.5, and 6.0~$\mu$m , and enstatite grains with sizes
0.1 and 1.5~$\mu$m (Dorschner et al. 1995; Servoin \& Piriou 1973; J\"{a}ger et al. 1998; 
Henning \& Mutschke 1997; Preibisch et al. 1993). 
We excluded from the fit the large enstatite grains (6.0~$\mu$m), 
since there is no evidence for them in our data. Moreover, their contribution 
resembles the continuum at 10~$\mu$m, and can mask gas features in the 13-14~$\mu$m 
region. We also excluded from the fit carbon grains, which had been previously 
considered by J09, since their contribution is only an additional
featureless continuum and does not affect the final silicate composition.
Note that the emission from large (6~$\mu$m) grains of most species differs little from the 
continuum emission (J09), so it is detectable only in objects with S/N$>$100-200. 
Therefore, the comparison of abundances and crystallinity
fractions is valid only for the grain sizes detected in each spectrum, which
depends on the S/N. 

Instead of assuming one or two given temperatures for the silicate and
continuum, the TLTD model considers that the silicate and the continuum 
originate in regions (disk inner rim, disk midplane, and optically thin disk atmosphere) 
that are characterized by a distribution of temperatures instead of a single one. 
These temperatures are parametrized as a power law.
The mass absorption coefficients are derived from
the material optical constants using the theory of Distribution of Hollow
Spheres for the crystalline dust (Min et al. 2005), and the classical Mie 
theory for spherical particles for the amorphous dust.  The flux then is calculated as:

\begin{eqnarray}
F_\nu = F_{\nu, {\rm cont}} & + & \sum_{i=1}^N\sum_{j=1}^MD_{i,j}\kappa_{i,j}
\int_{\rm{T_{\rm a, max}}}^{\rm{T_{\rm a, min}}}\frac{2\pi}{d^2}B_\nu(T){T}^{\frac{2-qa}{qa}}dT
\label{eq:1}
\end{eqnarray}

where the flux in the continuum is:

\begin{eqnarray}
F_{\nu, {\rm cont}} = D_0 \frac{\pi R_\star^2}{d^2} B_\nu(T_\star)&+& D1\int_{\rm{T_{\rm r,max}}}^{\rm{T_{\rm r, min}}}\frac{2\pi}{d^2}B_\nu(T){T}^{\frac{2-qr}{qr}}dT \\
&+& D2\int_{\rm{T_{\rm m,max}}}^{\rm{T_{\rm m, min}}}\frac{2\pi}{d^2}B_\nu(T){T}^{\frac{2-qm}{qm}}dT.
\label{eq:2}
\end{eqnarray}

Here, $R_*$ and $T_*$ are the radius and effective temperature of the
star, and $T_{a}$, $T_{r}$, and $T_{m}$ are the disk atmosphere, rim, and disk midplane
temperatures, which vary between a minimum and maximum value, and are power laws of the radius
(with exponents qa, qr, and qm, respectively). The exponents of the temperatures (qa, qr, qm) 
and the coefficients of each contribution ($D_0$, $D_1$, $D_2$, and $D_{i,j}$) are fitted
to the data using a genetic optimization algorithm (see details in J09).  The errors in the IRS spectra are taken 
into account by adding random Gaussian noise to the original spectrum at the noise level, and then
repeating the fit 100 times. The final silicate composition is obtained as the average of the whole 
set, and the asymmetric errors are derived from the standard deviation in both the positive
and negative direction, as was done for the Coronet cluster (SA08). 
The total crystalline fraction and the average grain sizes with their errors
are estimated in the same way. 

\begin{figure}
\plotone{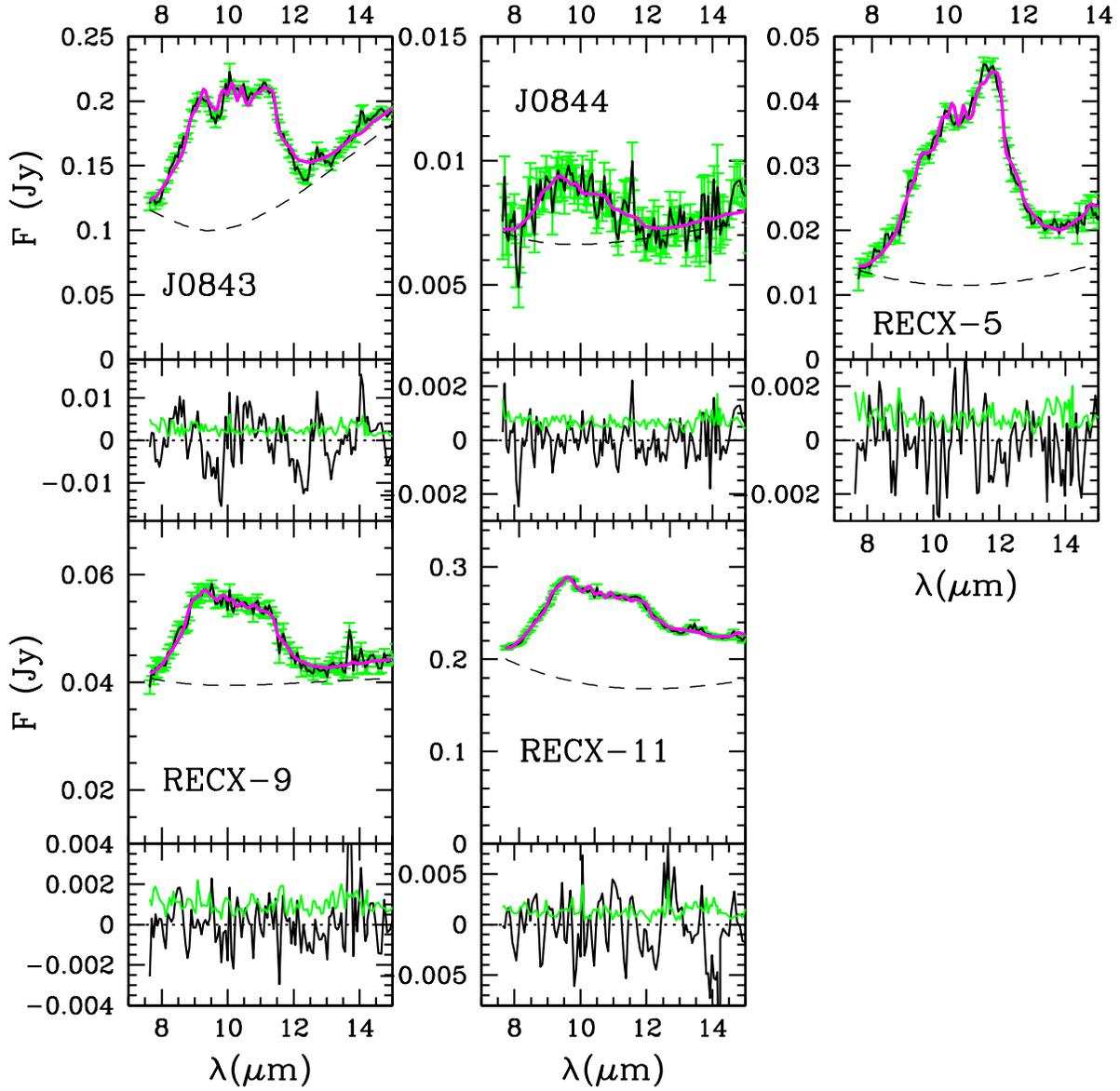}
\caption{Silicate fitting of the 10~$\mu$m feature of the stars in $\eta$ Cha,
using the TLTD method. The original spectrum is represented as a black solid line,
with the errors in green. The fit to the continuum is the dashed line, and the
model fit is the magenta thick solid line. The residuals (black), together with the
errors (green), are displayed under each source, and they do not reveal any systematic 
deviations. Note that the fit for J0844 only indicates
the presence of silicate emission, but cannot be used to obtain details about
mineralogy due to the low S/N of the spectrum.  \label{silfit-fig}}
\end{figure}

\begin{figure}
\plotone{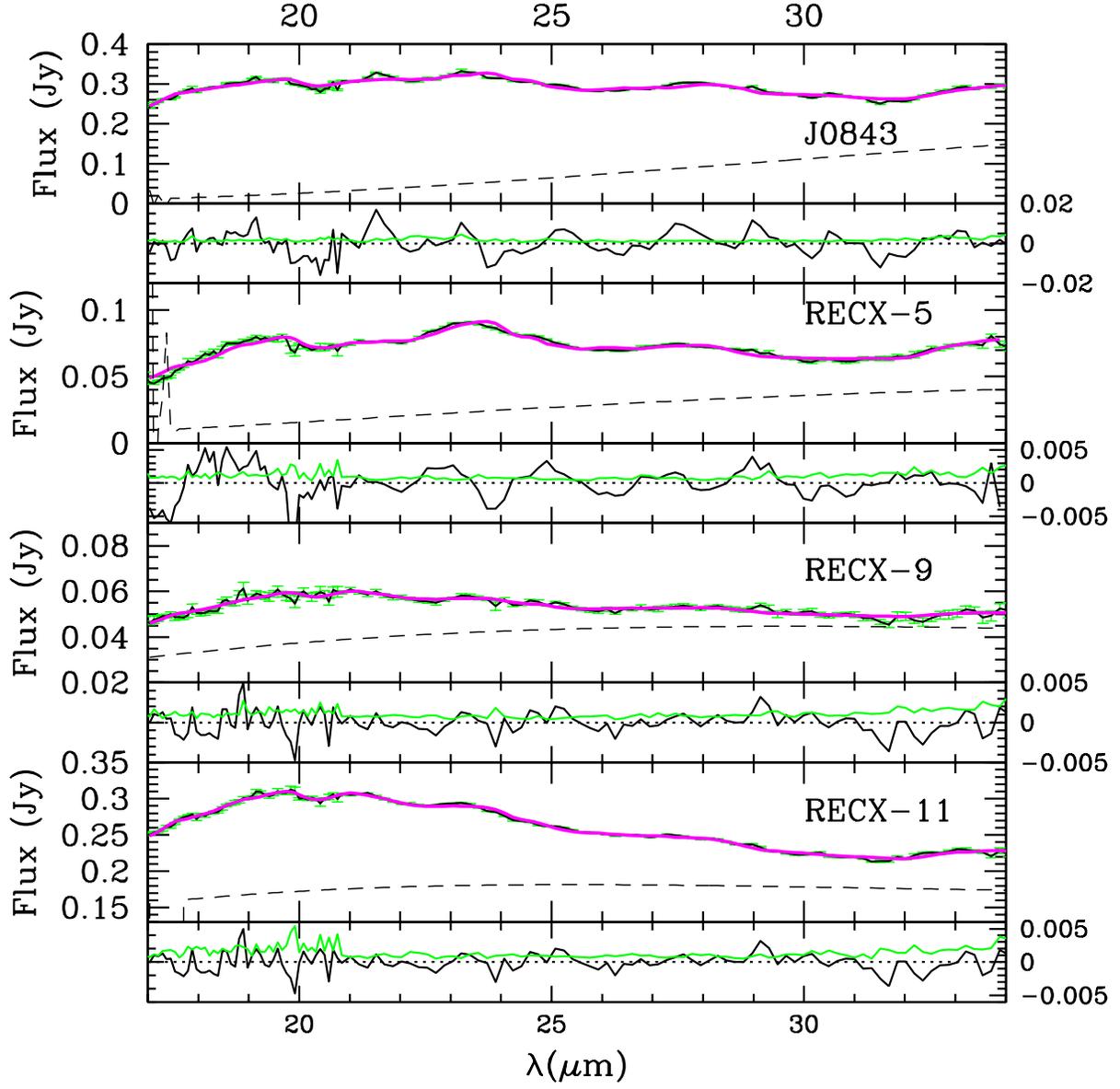}
\caption{Silicate fitting at 17-35~$\mu$m for the stars in $\eta$ Cha,
using the TLTD method. The original spectrum is represented as a black solid line,
with the errors in green. The fit to the continuum is the dashed line, and the
model fit is the magenta thick solid line. The residuals (black), together with the
errors (green), are shown in the panel below each object. \label{sillong-fig}}
\end{figure}

Tables \ref{sil-table},\ref{temp-table}, and \ref{sizecryst-table} list the outcome 
of the fits in terms of mass abundances of the different species, temperature of 
the silicate-producing region, average grain
sizes of the amorphous and crystalline silicates, and percentage of crystalline
grains versus total in the optically thin disk atmosphere. We only fitted the
objects with evidence of silicate emission. The applied spectral analysis method
assumes that the dust composition does not vary in the disk with radial distance 
(wavelength). Thus, in order to handle possible differences in radial composition, we
splitted the IRS spectra in two intervals (7-17~$\mu$m and 17-35~$\mu$m), which
were fitted separately. We fitted J0843, J0844, RECX-5, RECX-9, 
and RECX-11 in the 7-17~$\mu$m region, and all of them except J0844 for
the long-wavelength region. The fits are displayed
in Figures \ref{silfit-fig} and \ref{sillong-fig}, for the short and long wavelength 
ranges, respectively. The TLTD model and the material constants used are able to reproduce the 
observed features without any systematic deviation suggestive of other dust components
(see residuals in Figures \ref{silfit-fig} and \ref{sillong-fig}).

By fitting the 7-17~$\mu$m range, we find that the grain sizes in the optically 
thin disk atmospheres of the $\eta$ Cha cluster members are in all cases very 
different from what is seen in the ISM. The average grain size is $\sim$1-3~$\mu$m, 
and there is a significant fraction of crystalline silicates. In the 10~$\mu$m region, 
the crystalline fractions observed range between $\sim$30\% for J0843 to $\sim$7\% 
for RECX-11. In general, the crystalline fraction is higher than observed around
solar-type stars (Sicilia-Aguilar et al. 2007), and comparable to those observed
in low-mass stars and BD (Apai et al. 2005; SA08). 
In low-mass objects, the 10~$\mu$m feature samples a hotter, closer-in region than
in solar-type stars. This may, at least in part, be the cause of these differences,
if crystalline silicates are produced in the innermost disk. Large 
crystalline grains are rare (probably because they form large aggregates together
with amorphous grains; Bouwman et al. 2008), and the emission from large amorphous grains 
($>$6-10~$\mu$m) resembles the continuum at 10~$\mu$m. Therefore, grain growth 
can also increase the measured crystalline fraction by ``hiding'' the amorphous
dust into few-micron-sized ($>$6~$\mu$m) grains. Moreover, large grains can
be removed from the disk atmosphere with time as settling increases (Sicilia-Aguilar
et al. 2007), so the proportions between small crystalline grains and large
amorphous grains can vary with age. Strong grain growth (lack of grains $<$6~$\mu$m), 
resulting in the absence of the 10~$\mu$m silicate feature in optically thick, accreting 
disks has been observed for early M stars in intermediate-aged
regions (Sicilia-Aguilar et al. 2007). The 
differences in the processes producing crystalline silicates and in the regions where
crystallinization occurs in solar-type and low-mass stars could also affect the
crystalline fractions observed, as well as the properties of the crystals. In addition,
if the disks of low- and very low-mass objects show structural differences (Hartmann
et al. 2006), also the radial transport of processed grains in the disk could be affected.

The silicate features in the 17-35~$\mu$m region is sensitive to colder dust 
and thus traces larger radial distances than the 10~$\mu$m
feature. Unless grain processing and growth occur at a similar rate at all radii and/or
radial mixing is very efficient, we would  expect to observe a radial
variation of the grain composition. For the objects with good S/N 
in the 17-35~$\mu$m region (J0843, RECX-5, RECX-9, RECX-11;
see Table \ref{sil-table}), the differences in grain sizes between both regions
are not significant, except for RECX-11, which shows larger amorphous grains in the 10~$\mu$m
region than at longer wavelengths. Nevertheless, the crystalline fractions
are in general different, with a tendency to larger crystalline fractions at
shorter wavelengths (closer-in radial distances). Longward of 17~$\mu$m it is hard
to estimate the total mass of amorphous silicate grains, since they lack sharp
features in this range. Therefore, the variations in crystallinity with radius
must be handled with care. For RECX-5 we observe differences
in the crystalline silicate composition with radius: The enstatite to forsterite
ratio decreases with wavelength (i.e., decreases with radial distance).
This is the opposite effect we would expect from the dust coagulation model
by Gail (1998, 2004), and has been observed in other regions (Bouwman et al. 2008;
Meeus et al. 2009), which suggests that the conversion from forsterite to enstatite in 
chemical equilibrium does not take place in these disks. 

Despite the limitations of small number statistics, 
the differences in silicate composition with the radial distance
point to the importance of high temperatures in
dust processing, and suggest a not very efficient radial mixing throughout the disk.
We must nevertheless remember that old clusters have usually very small disk
fractions. Our results for the $\eta$ Cha cluster are complete in the sense
that they include all the known cluster members, which have been studied
in detail. Nevertheless, they comprise a too small population to extrapolate
these results to other regions with similar ages.
Extending these observations to a larger number of objects and to regions with 
different ages would be needed to study the variations in grain processing
and radial mixing with the evolutionary stage of the disk and the age of the system.

\subsection{Dust evolution? The $\eta$ Cha cluster and its younger ``twin'', the Coronet cluster}

As we mentioned before, the low-mass $\eta$ Cha cluster strongly
resembles the young ($\sim$1 Myr) Coronet cluster.
Comparing $\eta$ Cha to the Coronet cluster, we observe very
similar mineralogy and crystalline fractions for the low-mass members 
that display silicate features (SA08), confirming
that the dust composition derived from the silicate features cannot be used 
as a proxy for evolutionary stage nor age. This is probably due to the fact 
that the silicate feature traces only a minimal part of the dust content 
in the disk upper layers. It also suggests that the first grain processing (growth
to few-$\mu$m sizes and crystallization) occurs at a very early stage in the disk. 
The typical spectral types of the cluster members are similar in both 
cases, although the Coronet cluster harbors two objects near or at the Hydrogen 
burning limit. 

Although the derived mineralogy is similar for the $\eta$ Cha and the Coronet
cluster members, we observe other evolutive signatures when we compare their IRS
spectra. About half of the disks in the Coronet cluster do not show 
10~$\mu$m features, despite having an excess in the 10~$\mu$m region and being optically 
thick, protoplanetary disks (taking into account their IR excess over the photosphere). 
In the $\eta$ Cha cluster, 5 out of the 6 optically thick disks around the low-mass
objects show silicate emission at 10~$\mu$m. The lack of silicate emission 
has been considered as a sign of evolution via grain growth, as large 
($>$6-10~$\mu$m) grains produce only continuum emission in the 10~$\mu$m region. 
Nevertheless, a counterintuitive correlation between the size of the grains and the 
age (with older objects showing more pristine silicate features produced by small 
amorphous grains) has been noticed among solar-type stars in the Cep OB2 
region (Sicilia-Aguilar et al. 2007). 
This anticorrelation between grain size and age seems related to the stronger
settling and lower turbulence levels/accretion rates expected at older ages,
which would not allow large grains to remain in the upper disk atmosphere 
(Sicilia-Aguilar et al. 2007). The correlation between grain size and accretion rate 
observed in Cep OB2 (Sicilia-Aguilar et al. 2007) and the large grains detected in FU Ori 
objects (Quanz et al. 2007) points in the same direction.
The comparison between the $\eta$ Cha cluster 
and the young Coronet cluster reveals now the same effect around low-mass (M-type) 
objects. From these observations, we can conclude that
large grains would form very early in the disk lifetime (as suggested by 
Dullemond \& Dominik 2005), and the grain population in the disk atmosphere
probably depends on the effect of turbulence in the dust settling. 

Accretion rates in the $\eta$ Cha cluster are low 
(Lawson et al. 2004; Table \ref{optlit-table}).
Although there are no measurements of the accretion rates in the Coronet cluster,
the H$\alpha$ observations and line profiles suggest higher accretion rates than 
for the $\eta$ Cha members and thus, more turbulent disks (SA08). This is
also the general trend we would expect from the studies of accretion rate evolution
versus age (Muzerolle et al. 2000; Sicilia-Aguilar et al. 2006b), given the $\sim$7 Myr
age difference between both regions. Within the $\eta$ Cha cluster, 
we do not see any significant correlation between the grain size and the accretion rate
(Lawson et al. 2004), although there are only 4 objects with known accretion rates, 
which are all very small (see Table \ref{star-table}). The way
settling and turbulence affect the the vertical grain distribution, the composition
of the disk atmosphere, and the appearance of the silicate feature is not clear
(Dullemond \& Dominik 2008). Therefore, this trend should be further explored and 
contrasted with other stellar and environmental properties. 
Other parameters affecting the disk structure, like differences in stellar mass/spectral type, or 
differences in the disk mass, evolution, and environment, could be partially
responsible for this effect.  

Taking into account the IR photometry of the $\eta$ Cha members,
it had been suggested that these disks would have suffered strong
grain growth (Haisch et al. 2005; Megeath et al. 2005). While this
is probably true, and may be responsible for the flattened SEDs and TO observed 
for the $\eta$ Cha members (see Section \ref{evol}), it cannot be 
confirmed using the silicate emission at IR wavelengths. The large grains 
are most likely larger than a few $\mu$m, and have probably settled
to the disk midplane, so they are not present in the disk atmosphere, where
the optically thin emission arises. In addition, IR observations are not sensitive to
the bulk of the dust content in the disk, including
the potential very large grains (more than a few $\mu$m in size, maybe even pebble-sized)
that are likely to have formed in the 8 Myr life of the disks around $\eta$ Cha members.
Observations at millimeter and sub-millimeter wavelengths are needed to trace
the presence and mass distribution of larger grains in these disks.

\subsection{$\eta$ Cha: A young debris-disk rather than a classical Be star \label{etacha}}

The B9 star $\eta$ Cha is the brightest member of the cluster that bears its name.
Its fast rotation, together with the presence of a flattened, featureless 
disk, suggested that it could be a classical Be star (see the review by
Porter \& Rivinius 2003). Traditionally, classical Be stars had been considered 
as an older stage in the evolution of very fast rotating B and early A stars. 
These stars would lose material at their equator, producing a gaseous hot disk that is 
bright in the IR via free-free emission (Gehrz et al. 1974). Initial surveys of young 
clusters suggested that there were no classical Be stars in regions aged
$<$10 Myr (Capilla et al. 2000). Nevertheless, more recent studies revealed
several classical Be stars in young regions (Bhavya et al. 2008). Even the $\sim$4 Myr old
cluster Tr 37 possesses a classical Be star, KUN-314s  (Sicilia-Aguilar et al. 2007). 
Aged $\sim$8 Myr, $\eta$ Cha would be among the youngest Be stars known.
Nevertheless, it could as well be a young debris disk, given that the 
excess emission over the stellar photosphere is comparable to what is seen in 
young debris disks a few Myr after planet formation 
(Kenyon \& Bromley 2005). In addition, there are known debris disks with 
mid-IR SED slopes similar to $\eta$ Cha (Sicilia-Aguilar et al. 2007).

To test both possibilities, we have compared the optical spectrum of 
$\eta$ Cha taken with UVES (from the ESO data archive) to the HIRES spectra of 
two young B stars in Tr 37 (kindly provided by L. Hillenbrand), the classical Be 
star KUN-314s and the B7 star MVA-437
(Contreras et al. 2002; Sicilia-Aguilar et al. 2005, 2007).
$\eta$ Cha was observed with UVES under the ESO program 
ID 66.D-0284(A) in 2001-02-18. All the data 
were obtained with the DICHROIC\#2 mode, which has two wavelength settings, 
BLUE (305-385~nm) and RED (575-945~nm, split up in two wavelength ranges 
REDL and REDU). The 8 spectra (4 taken with the BLUE setting, and 4 taken with the
RED setting) were reduced  using the standard UVES pipeline. 
Integration times were short, 5-15~s for the red part, and 10~s for the blue, as $\eta$ Cha is 
extremely bright for UVES mounted on a UT. The final spectrum is the sum of these spectra.
The HIRES spectra of KUN-314s and MVA-437
were taken in 2001, before the HIRES upgrade, using the rg610 filter, which
provides a wavelength coverage between 6350 and 8750\AA\ with a resolution $\sim$34,000.
The spectra were reduced using the MAKEE Keck Observatory HIRES data reduction software,
developed by T. Barlow. 

\begin{figure}
\plottwo{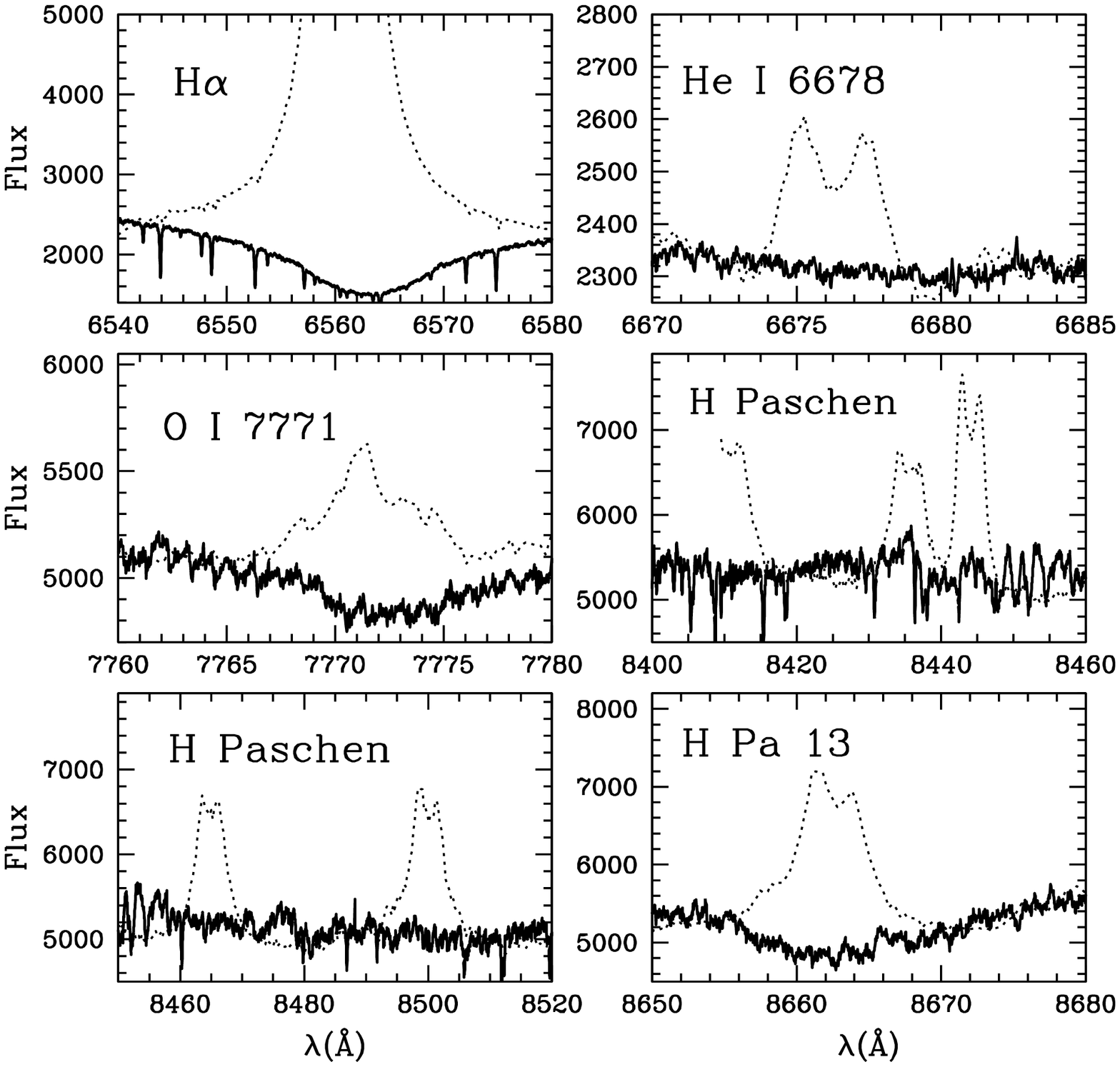}{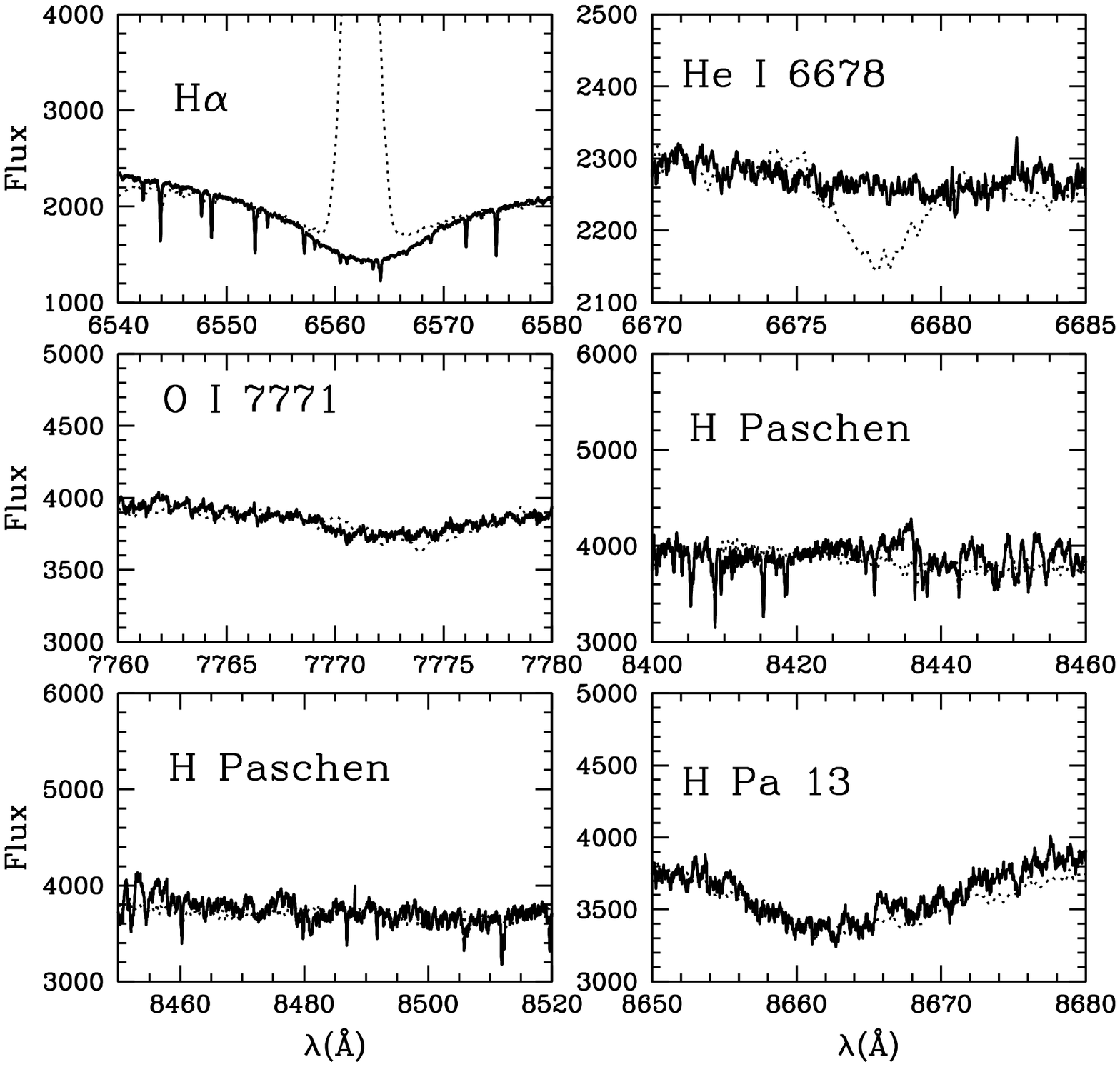}
\caption{Comparison of the optical spectrum of $\eta$ Cha (thick line) with the classical Be
star KUN-314s (left, dotted line) and the B7 star with a transition/debris disk MVA 437 (right, 
dotted line), both young ($\sim$4 Myr) stars in Tr 37 (Sicilia-Aguilar et al. 2007). 
The spectra have been scaled to match the continuum levels. The $\eta$ Cha spectrum is fully consistent
with a normal young B star, with some variations due to rotation, accretion/activity (MVA 437 presents
H$\alpha$ emission due to either chromospheric activity or very low accretion, while $\eta$ Cha
 does not) and spectral type. The flux is given in arbitrary units.\label{etachaBe-fig}}
\end{figure}

The optical spectra were scaled to match their continuum levels and compared 
to determine the presence of the emission lines typical of classical Be stars 
(Figure \ref{etachaBe-fig}). The comparison reveals
that $\eta$ Cha does not show any of the emission lines typical of 
classical Be stars like KUN-314s. In particular, it lacks the H and He emission
lines with the double-peaked rotation profiles characteristic of classical Be stars.
On the contrary, the spectrum of $\eta$ Cha is very similar to that of the
B7 star MVA-437, except for the accretion-related 
H$\alpha$ line, for which MVA-437 shows narrow emission but the 
non-accreting $\eta$ Cha displays only rotationally broadened absorption. 
Some differences 
in the absorption lines (line broadening, lack of He absorption lines in $\eta$ Cha)
are due to differences in rotation and small differences in the 
spectral type, since $\eta$ Cha is a faster rotator than MVA-437 and a B9 star,
instead of a B7 star like MVA-437. 

\begin{figure}
\plottwo{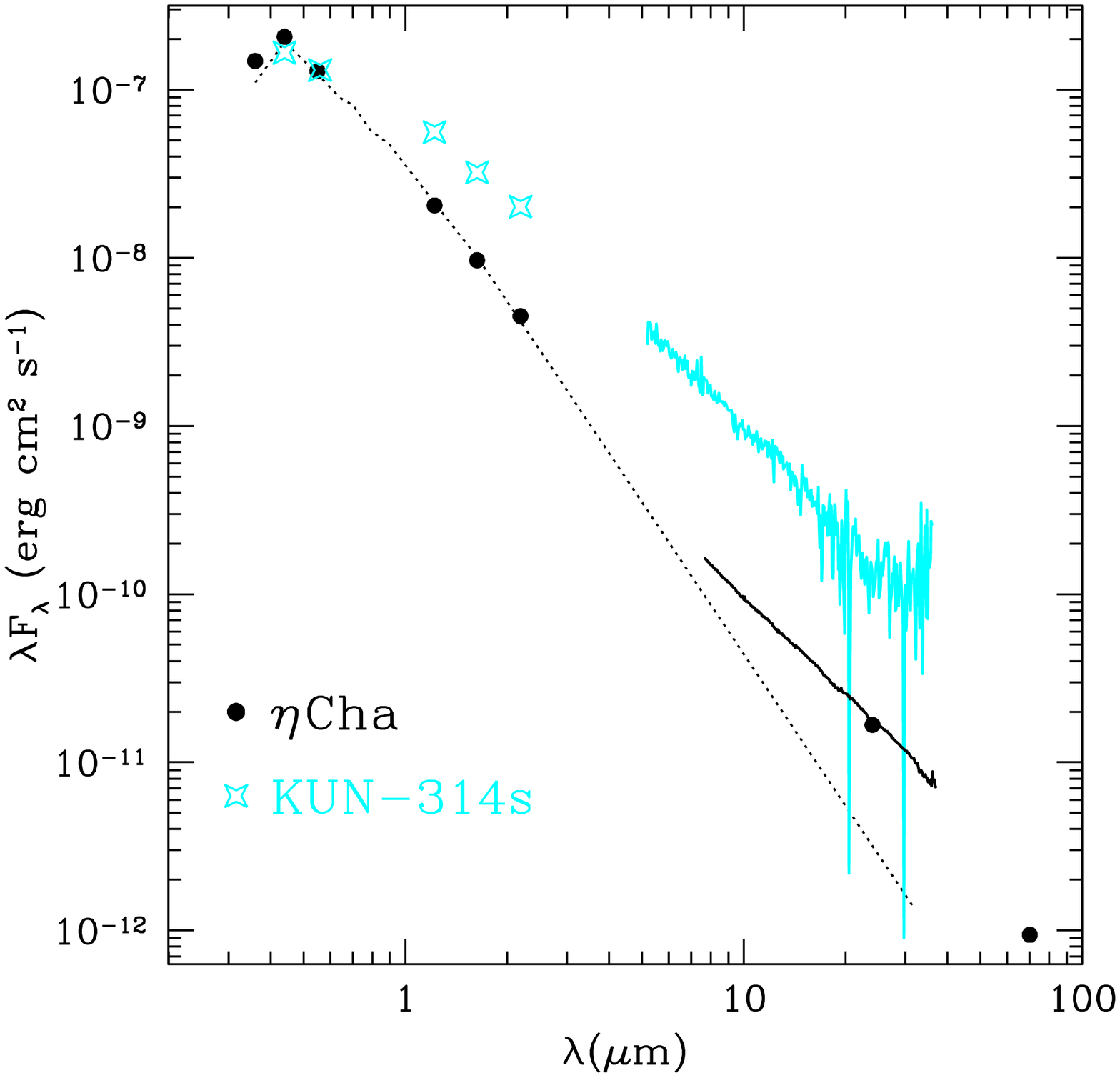}{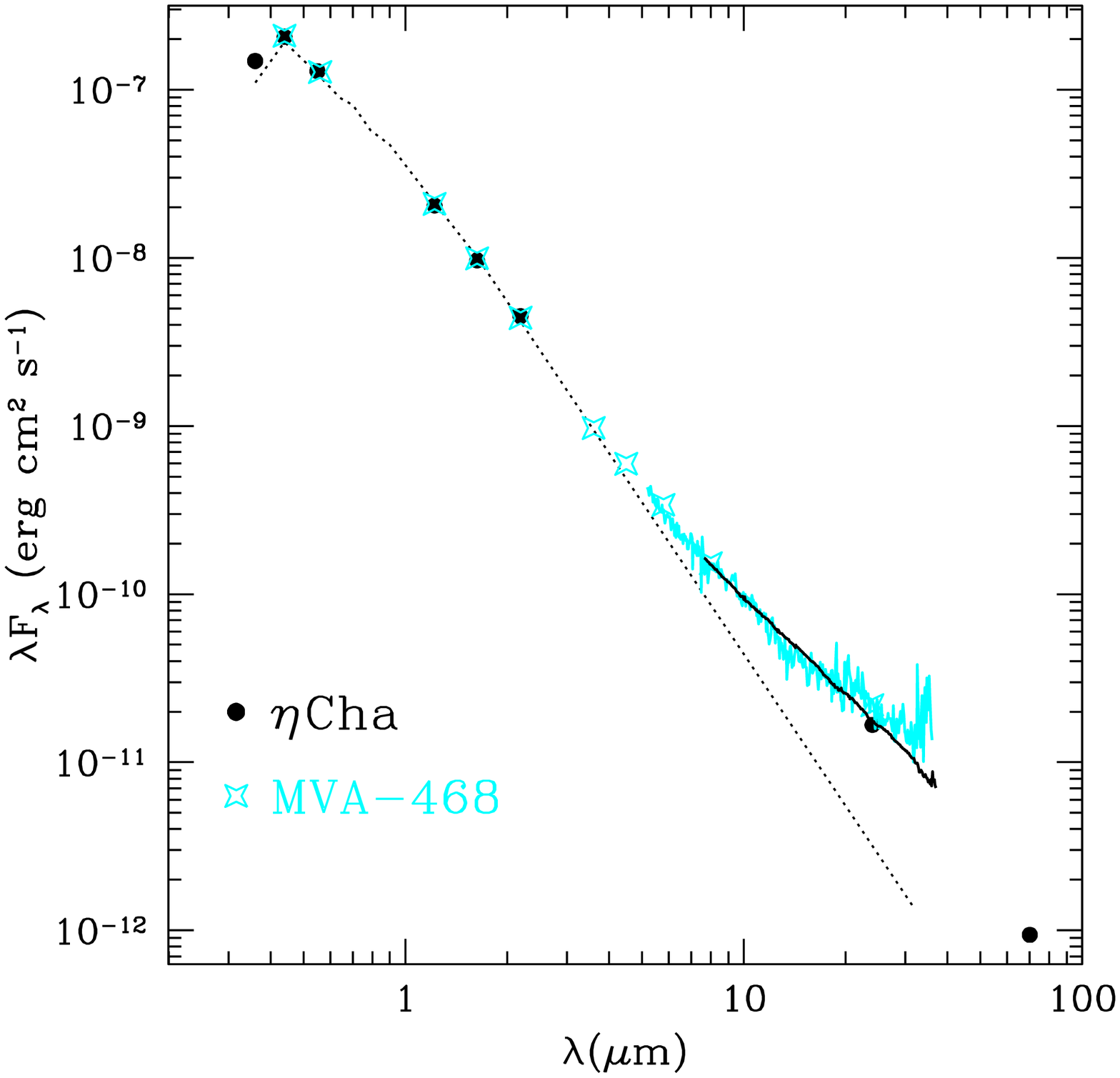}
\caption{Comparison of the SEDs of $\eta$ Cha with the classical Be star
KUN-314s (left) and the B7 star with a transition/debris disk MVA 468 
(right; Sicilia-Aguilar et al. 2007). The SEDs of KUN-314s and MVA-468
have been scaled to their optical and JHK fluxes to match the $\eta$ Cha 
distance. Although the slope of the $\eta$ Cha disk is roughly consistent 
with that of a classical Be star, as it lacks features, those 
characteristics are not unique of classical Be stars, being present in
some transitional and/or young debris disks. \label{etachaSED-fig}}
\end{figure}

Another reason that suggested that $\eta$ Cha could be a classical Be star was the
slope of its SED, similar to the expected slope for free-free emission 
found in classical Be stars (Gehrz et al. 1974). KUN-314s is a late-B 
star (the uncertain spectral type is due to the double-peaked emission lines masking most 
of the A and B spectral type features), having thus a spectral 
type similar to $\eta$ Cha. Therefore,
we would expect a very similar emission from a classical Be ``decretion''
disk if both objects have the same nature, with some variations depending on
whether the decretion disk is optically thin or optically thick (Gehrz et al. 1974). 
For comparison, the SED of KUN-314s was scaled to match the distance and luminosity of
$\eta$ Cha (Figure \ref{etachaSED-fig}).
The wavelength at which the IR excess starts is different in $\eta$ Cha
and in KUN-314s. In fact, the entire $\eta$ Cha SED is more similar to
the debris disk around the B7 star MVA-468. Both $\eta$ Cha
and MVA-468 lack silicate and gas features, and have a slope similar to that
expected from free-free emission, although the 70~$\mu$m flux in $\eta$ Cha
suggests a deviation from the free-free emission slope at long wavelengths. 
Debris and ``transitional''
disks with such slopes have been found not only around B-type stars, but also
around F-type objects (e.g., MVA-447 in Tr 37; Sicilia-Aguilar et al. 2007).
Although some of the debris and/or ``transition'' disks around intermediate-mass
stars show silicate features
in the 10-30~$\mu$m region, most of them display flat, featureless spectra 
(Sicilia-Aguilar et al. 2007), probably because of the lack of small ($<$10~$\mu$m)
dust grains. Due to its high luminosity, the silicate feature in a B9 star 
is produced by small grains located at up to $\sim$70-100 AU distance, so
the total absence of silicate emission in the IRS spectrum points to the lack of 
small grains in most or even in the whole disk. Then, B stars with featureless
mid-IR spectra contain probably second-generation, collisional, reprocessed debris 
disks made of larger grains (Kenyon \& Bromley 2005).

Therefore, we conclude that there is no evidence of $\eta$ Cha being a
classical Be star. Despite its fast rotation, it is not significantly
different from a normal, young B star surrounded by a disk with a
large inner hole. The disk could be either the leftover from an evolved 
protoplanetary disk that has been 
evacuated in its innermost regions by a non-detected companion, grain growth, 
photoevaporation, and/or planet formation (``transition'' disk), or more likely,
due to the lack of small dust grains, a debris disk made by reprocessed dust from 
collisions a few Myr after planet formation.

\section{Conclusions \label{conclu}}

The IRS spectra of the $\eta$ Cha cluster members presented here cover probably
the largest and best studied sample of disks in a $\sim$8 Myr old cluster, and
are particularly interesting to understand disk evolution, given that the
stellar properties of the $\eta$ Cha members (spectral types, ages, accretion,
and in most cases, binarity) are well known. Moreover, the $\eta$ Cha cluster
is a key to understand disk evolution, as it still contains a large population of
disks for its old age.

The IRS data confirms a relatively high disk fraction for
a $\sim$8 Myr-old cluster, $\sim$50\%, among the low-mass stars (spectral types
late K-M). Half of the disks are TO, which is a higher
fraction than observed for solar-type stars in other regions (e.g., Taurus, 
Tr 37, Orion), even if we consider the effects of small number statistics
and the differences in luminosity between M-type and solar-type stars.
The fraction of TO is comparable to the TO fraction in the young ($\sim$1 Myr), 
low-mass star Coronet cluster and in MBM12
($\sim$2 Myr). The large number of TO (including flattened, ``homologously
depleted'' TO) in clusters with very different 
ages suggest that at least some TO are long-lived, and therefore, not in an 
intermediate, rapid ``transitional'' stage between Class II and Class III objects.

Disks around low-mass stars have lower masses than the disks around solar-type
TTS, so they might also have different structure (Hartmann et al. 2006) and/or 
evolve along a different path than their solar-type companions.
Long-lived ``transition'' objects do not fit within the classical picture of rapid disk 
dissipation, suggesting that they could be part of the ``normal'' type
of disk around very low-mass objects. The presence of long-lived inner holes 
and depleted inner disks could be
related to planetesimal or even planet formation, non-resolved binary 
companions, or be the result of the initial conditions (disk mass and structure)
and formation environment for these low-mass objects. 
The TO in $\eta$ Cha have zero or very low accretion rates, which could
be compatible with the planet formation hypothesis. In addition, the fact that
one of the Class II disks has a very low accretion rate (RECX-11) and no inner
hole suggests that photoevaporation, expected to remove disks in
timescales $\sim$10$^5$ years, may not be as efficient to open inner holes as
previously thought.
Further observations,
especially those focused on the detection of close-in companions, are
now required to study the origin and properties of accreting and non-accreting
TO.

The IRS spectra of the low-mass cluster members also reveal that the 
dust grains in the optically thin disk atmospheres have suffered substantial grain 
processing, being very different from ISM grains.
They are typically large (1-3~$\mu$m) and contain
different proportions of crystalline silicates ($\sim$10-30\%) and a large variety
of grain compositions. The differences
between the 7-17~$\mu$m and the 17-35~$\mu$m features show differences in grain
composition and solid state with the radial distance, suggesting inefficient
radial mixing, and the importance of high temperatures in dust processing.

The diversity of cases, and the fact that the grains
are not different from what is found in objects with similar spectral types
in younger clusters (e.g. the Coronet cluster), suggest that the grain 
size/crystalline fraction observed cannot be used as a proxy for age nor evolutionary 
state. Few-$\mu$m grains and crystalline silicates must then form very early in
the disk life. The grain population in the disk atmosphere
is probably more affected by the turbulence levels in the disk than by
the global grain growth. Therefore, objects with SEDs lacking silicate features, having 
presumably disk atmospheres dominated by large grains, appear to be more
common in younger, more turbulent disks. IR observations are not sensitive
to grains larger than $\sim$10~$\mu$m, whose mass fractions and sizes might
be better tracers of the evolutionary state. 

The combination of IRS, MIPS, and UVES data on the B9 star $\eta$ Cha
that gives the name to the cluster reveals that it is surrounded 
by a debris disk or a very evolved protoplanetary disk with a large inner hole and
no small dust grains, rather than being a young, classical Be star as it had
been suggested. The luminosity of this disk is consistent with the predictions
for a second-generation debris disk a few Myr after planet formation (Kenyon \& Bromley
2005).

Finally, we make available the corrected MIPS photometry for the $\eta$ Cha cluster
members, since the data given in the literature was probably affected by 
errors in the data reduction process and/or in the photometry. 
Our revised MIPS photometry is fully  consistent with the fluxes in the IRS data.

We want to thank L. Hillenbrand for providing the HIRES spectra
of the intermediate-mass members of Tr 37, KUN-314s and MVA-437.
We also thank the anonymous referee for his/her review, interesting
suggestions and useful comments, which have greatly contributed to the clarity of 
this paper. 
A.S-A. acknowledges support from the Deutsche Forschungsgemeinschaft, DFG,
grant number SI 1486/1-1.
This work is based on observations made with the {\it Spitzer\,} Space Telescope, 
which is operated
by the Jet Propulsion Laboratory, California Institute of Technology under a contract with 
NASA. It also makes use of data products from the Two Micron All Sky Survey, which is 
a joint project of the University of Massachusetts and the Infrared Processing and Analysis 
Center/California Institute of Technology, funded by the National Aeronautics and Space 
Administration and the National Science Foundation.

\clearpage

\clearpage

%\clearpage

\clearpage

\clearpage

\clearpage

\clearpage

\clearpage

\begin{deluxetable}{lccccccc}
\tabletypesize{\scriptsize}
%\rotate
\tablenum{1}
\tablecolumns{8} 
\tablewidth{0pc} 
\tablecaption{Stellar parameters\label{star-table}} 
\tablehead{
 \colhead{Name} & \colhead{Sp. Type} & \colhead{H$\alpha$ EW (\AA)} & \colhead{\.{M} (10$^{-8}$ M$_\odot$yr$^{-1}$)} & \colhead{H$\alpha$ Type} & \colhead{SED Type}  & \colhead{Binary} & \colhead{References}} 
\startdata
$\eta$ Cha &  B8+M1:$^a$	& 8.5	&	&  	& Class III/debris  &    	 & 1	\\
RS Cha 	&  A7+A8+M1:$^a$	& 9.2	&	& 	& Class III  & eclipsing A7/A8 & 1	\\
HD 75505 & A5		&	&	&	& Class III  &  	& 2	\\
J0836	& M5.3+M5.3:	&	&	&	& Class III  & $<$0.04 	& 3	\\
J0838	& M5.0+M5.0:	&	&	&	& Class III  & $<$0.04 	& 3	\\
J0841	& M4.7		& -12.0	&	& WTTS	& TO/flat    &          & 3,4	\\
J0843 	& M3.4		& -90	& 0.1	& CTTS	& Class II   &    	& 3,5	\\
J0844	& M5.5		&	&	&	& flat   &          & 3	\\
RECX-1 	& K7.0+M0.0	& -1.4	&	& WTTS	& Class III  &    0.1	& 3,1,6	\\
RECX-3 	& M3.0		& -2.2	&	& WTTS	& TO   	     &    	& 3,5	\\
RECX-4 	& M1.3		& -2.3	&	& WTTS	& TO 	     &   	& 3,5	\\
RECX-5 	& M3.8		& -35	& 0.005	& CTTS	& TO         &   	& 3,5	\\
RECX-6 	& M3.0		& -3.6	&	& WTTS	& Class III  &    	& 3,5	\\
RECX-7 	& K6.9+M1.0	& -0.4	&	& WTTS	& Class III  &   0.001	& 3,5,7,8	\\
RECX-9 	& M4.4+M4.7	& -10	& 0.004	& CTTS	& TO    &   0.2    & 3,5,6	\\
RECX-10 & M0.3		& -1.0	&	& WTTS	& Class III  &    	& 3,5	\\
RECX-11	& K6.5		& -3.2	& 0.004	& CTTS	& Class II   &   	& 3,5	\\
RECX-12	& M3.2+M3.2:	& -4.2	&	& WTTS	& Class III  &   0.04 	& 1,3,5,9	\\
\enddata
\tablecomments{Uncertain values are marked with `:'. 
The SEDs are classified as Class II (comparable to Taurus),
flattened disks ($\lambda$F$_\lambda \sim \lambda^{-4/3}$),
TO (zero or very small near-IR excess), and Class III
(no IR excess over the stellar photosphere), see text for
more details. $\eta$ Cha is defined as a Class III source with
a second generation ``debris'' disk. The ``binary'' column 
indicates the projected separation (in arcsec), when
known, the spectral types of the two components are listed in
the corresponding column. $^a$: The low-mass companions to 
$\eta$ Cha and RS Cha are speculative, based on the X-ray
emission from these stars having roughly the properties of
a roughly 0.5 solar mass T Tauri star (spectral type $\sim$M1).
References for the spectral types, H$\alpha$, and accretion
status: 1 = Mamajek et al. (1999);
2 = Lawson et al. (2001);
3 = Lyo et al. (2004b); 
4 = Lawson et al. (2002);
5 = Lawson et al. (2004);
6 = K\"{o}hler \& Petr-Gotzens (2002);
7 = Moraux et al. (2007);
8 = Lyo et al. (2003);
9 = Brandecker et al. (2006).
Note that for the H$\alpha$ class, we give priority to the detection of
broad H$\alpha$ lines (sign of accretion) to classify an object as a CTTS,
rather than to the H$\alpha$ EW alone, that may be misleading in cases
of very low mass accretion rates (Sicilia-Aguilar et al. 2006b).
}
\end{deluxetable}

\clearpage

\begin{deluxetable}{llcccccl}
\tabletypesize{\scriptsize}
%\rotate
\tablenum{2}
\tablecolumns{8} 
\tablewidth{0pc} 
\tablecaption{Optical photometry from the literature\label{optlit-table}} 
\tablehead{
 \colhead{Name} & \colhead{2MASS ID} &\colhead{U-B}& \colhead{B-V}& \colhead{V}& \colhead{V-R$_c$}& \colhead{V-I$_c$} & \colhead{Refs.}} 
\startdata
$\eta$ Cha &  08411947-7857481 & -0.35 & -0.10 & 5.47 & --- & --- & 1,2 \\
RS Cha &  08431222-7904123 & 0.07 & 0.24 & 6.04 & --- & --- &  2,3  \\
HD-75505  & 08414471-7902531 & --- & --- & 7.27 & 0.06 & 0.15 &  2,5,6  \\
J0838 &  08385150-7916136 & --- & --- & 16.82 & 1.61 & 3.54 &   7,2\\
J0841 &  08413030-7853064 & --- & --- & 17.07 & 1.52 & 3.25 &  7,2\\
J0843 &  08431857-7905181 & --- & --- & 13.97 & 0.99 & 2.20 &  2,7 \\
J0844 &  08440914-7833457 & --- & 0.49 & 17.74 & --- & --- &   2,8 \\ 
J0836 &  08361072-7908183 & --- & --- & 17.66 & 1.72 & 3.76 & 2,6 \\
RECX-1  & 08365623-7856454 & --- & 1.23 & 10.61 & 0.71 & 1.42 & 2,6\\
RECX-3  & 08413703-7903304 & --- & 1.51 & 14.37 & 1.16 & 2.58 & 2,4,5 \\
RECX-4  & 08422372-7904030 & --- & 1.45 & 12.73 & 0.93 & 1.92 & 2,4,5\\
RECX-5  & 08422710-7857479 & --- & 1.51 & 15.20 & 1.29 & 2.81 &   2,4,5\\
RECX-6  & 08423879-7854427 & --- & 1.50 & 14.08 & 1.00 & 2.40 &   2,4,5\\
RECX-7  & 08430723-7904524 & --- & 1.19 & 10.84 & 0.67 & 1.40 &  2,4,5\\
RECX-9  & 08441637-7859080 & --- & 1.52 & 15.00 & 1.36 & 2.99 &  2,4,5 \\
RECX-10  & 08443188-7846311 & --- & 1.45 & 12.53 & 0.86 & 1.78 &   2,4,5\\
RECX-11  & 08470165-7859345 & --- & 1.22 & 11.13 & 0.66 & 1.37 & 2,4,5 \\
RECX-12  & 08475676-7854532 & --- & 1.51 & 13.17 & 1.07 & 2.37 &  2,4,5 \\
\enddata
\tablecomments{Photometry observations of the $\eta$ Cha members.
References: 1 = Johnson et al. (1966); 2 = 2MASS, Cutri et al. (2003);
3 = Mermilliod (1991); 
4 = Lawson et al. (2001); 5 = Lyo et al. (2004); 6 = Lyo et al. (2003);
7 = Lawson et al. (2002); 8 = NOMAD Catalog, Zacharias et al. (2005),
provided by Vizier. }
\end{deluxetable}

\clearpage

\begin{deluxetable}{llcccccccccc}
\tabletypesize{\scriptsize}
\rotate
\tablenum{3}
\tablecolumns{11} 
\tablewidth{0pc} 
\tablecaption{IR photometry\label{IR-table}} 
\tablehead{
 \colhead{Name} & \colhead{2MASS ID} &\colhead{J}& \colhead{H}& \colhead{K}& \colhead{3.6$\mu$m}& \colhead{4.5$\mu$m} &\colhead{5.8$\mu$m} & \colhead{8.0$\mu$m} & \colhead{24$\mu$m} & \colhead{70$\mu$m}} 
\startdata
$\eta$-Cha & 08411947-7857481 & 5.69$\pm$0.02 & 5.72$\pm$0.04 & 5.72$\pm$0.02 & $^a$ & $^a$ & $^a$ & $^a$  & 4.33$\pm$0.02 & 3.87$\pm$0.13:  \\
RS-Cha & 08431222-7904123 &  5.99$\pm$0.03 & 5.88$\pm$0.04 & 5.85$\pm$0.03 & $^a$ & $^a$ & 5.40$\pm$0.01 & 5.40$\pm$0.01  & 5.34$\pm$0.03 & ---  \\
HD-75505 & 08414471-7902531 &  7.06$\pm$0.03 & 6.99$\pm$0.04 & 6.93$\pm$0.02 & 6.97$\pm$0.01 & 6.97$\pm$0.01 & 6.97$\pm$0.01 & 6.97$\pm$0.01  &  6.97$\pm$0.06 & ---  \\
J0836 & 08361072-7908183 & 11.85$\pm$0.02 & 11.28$\pm$0.03 & 10.95$\pm$0.02 & 10.57$\pm$0.03 & 10.45$\pm$0.01 & 10.40$\pm$0.01 & 10.37$\pm$0.01  & $^b$ & $^b$   \\
J0838 & 08385150-7916136 &  11.28$\pm$0.02 & 10.72$\pm$0.02 & 10.43$\pm$0.02 & 10.07$\pm$0.02 & 10.00$\pm$0.01 & 9.92$\pm$0.01 & 9.90$\pm$0.01  & $^b$ & $^b$ \\
J0841 & 08413030-7853064 &  11.81$\pm$0.03 & 11.24$\pm$0.03 & 10.98$\pm$0.02 & 10.53$\pm$0.01 & 10.26$\pm$0.05 & 10.04$\pm$0.01 & 9.48$\pm$0.01  &  7.07$\pm$0.06 & ---  \\
J0843 & 08431742-7905236 &  13.53$\pm$0.05 & 13.18$\pm$0.08 & 13.09$\pm$0.04 & 8.38$\pm$0.03 & 7.91$\pm$0.01 & 7.42$\pm$0.03 & 6.51$\pm$0.01  & 3.52$\pm$0.01 & 1.89$\pm$0.02  \\
J0844 & 08440914-7833457 & 12.51$\pm$0.02 & 11.98$\pm$0.02 & 11.62$\pm$0.02 & 11.02$\pm$0.01 & 10.75$\pm$0.01 & 10.42$\pm$0.01 & 9.76$\pm$0.01  &  7.11$\pm$0.07 & $^b$  \\
RECX-1 & 08365623-7856454 & 8.15$\pm$0.02 & 7.50$\pm$0.05 & 7.34$\pm$0.02 & 7.17$\pm$0.02 & 7.20$\pm$0.02 & 7.16$\pm$0.01 & 7.11$\pm$0.02  &  6.96$\pm$0.06 & ---  \\
RECX-3 & 08413703-7903304 &  10.35$\pm$0.02 & 9.65$\pm$0.02 & 9.41$\pm$0.02 & 9.27$\pm$0.02 & 9.21$\pm$0.02 & 9.09$\pm$0.05 & 9.15$\pm$0.01  & 8.49$\pm$0.13 & ---  \\
RECX-4 & 08422372-7904030 &  9.54$\pm$0.02 & 8.78$\pm$0.06 & 8.62$\pm$0.02 & 8.45$\pm$0.02 & 8.45$\pm$0.02 & 8.37$\pm$0.01 & 8.32$\pm$0.02  &  7.73$\pm$0.09 & ---  \\
RECX-5 & 08422710-7857479 &  10.78$\pm$0.02 & 10.10$\pm$0.02 & 9.86$\pm$0.02 & 9.59$\pm$0.03 & 9.50$\pm$0.03 & 9.37$\pm$0.01 & 8.89$\pm$0.01  &  5.05$\pm$0.02 & 2.38$\pm$0.04  \\
RECX-6 & 08423879-7854427 &  10.23$\pm$0.03 & 9.58$\pm$0.02 & 9.29$\pm$0.02 & 9.15$\pm$0.04 & 9.07$\pm$0.07 & 9.04$\pm$0.01 & 9.04$\pm$0.01  &  8.88$\pm$0.16 & ---  \\
RECX-7 & 08430723-7904524 &  8.42$\pm$0.02 & 7.76$\pm$0.03 & 7.63$\pm$0.03 & 7.52$\pm$0.01 & 7.54$\pm$0.02 & 7.49$\pm$0.01 & 7.46$\pm$0.01  &   7.37$\pm$0.08 & ---  \\
RECX-9 & 08441637-7859080 &  10.26$\pm$0.03 & 9.67$\pm$0.03 & 9.34$\pm$0.02 & 8.99$\pm$0.02 & 8.80$\pm$0.01 & 8.57$\pm$0.01 & 7.97$\pm$0.02  &  5.38$\pm$0.03 & 3.16$\pm$0.07  \\
RECX-10 & 08443188-7846311 &  9.65$\pm$0.02 & 8.92$\pm$0.06 & 8.73$\pm$0.02 & 8.58$\pm$0.01 & 8.61$\pm$0.01 & 8.55$\pm$0.02 & 8.49$\pm$0.01  &  8.45$\pm$0.13 & ---  \\
RECX-11 & 08470165-7859345 &  8.73$\pm$0.02 & 8.03$\pm$0.06 & 7.66$\pm$0.04 & 7.09$\pm$0.02 & 6.86$\pm$0.01 & 6.57$\pm$0.01 & 5.97$\pm$0.01  &  3.68$\pm$0.01 & 1.80$\pm$0.02  \\
RECX-12 & 08475676-7854532 &  9.32$\pm$0.02 & 8.68$\pm$0.08 & 8.41$\pm$0.03 & 8.19$\pm$0.01 & 8.15$\pm$0.01 & 8.10$\pm$0.02 & 8.07$\pm$0.01  & 7.95$\pm$0.10 & ---  \\
\enddata
\tablecomments{IR Photometry observations of the $\eta$ Cha members.
$^a$ = The object is saturated. $^b$ = The object
is out of the MIPS field. 
The JHK data is obtained from 2MASS (Cutri et al. 2003), and the IRAC data was
published in Megeath et al. (2005). The MIPS magnitudes have an extra error around
10\% due to conversion factors and different chip coverage.
Uncertain values are marked with `:'. }
\end{deluxetable}

\begin{deluxetable}{lccccc}
%\tabletypesize{\scriptsize}
%\rotate
\tablenum{4}
\tablecolumns{6} 
\tablewidth{0pc} 
\tablecaption{Silicate Composition in $\eta$ Cha (TLTD) \label{sil-table}} 
\tablehead{
 \colhead{Name} & \colhead{0.1$\mu$m A.Ol.}  & \colhead{0.1$\mu$m A.Py.}  & \colhead{0.1$\mu$m Fors.}  & \colhead{0.1$\mu$m Enst.}  & \colhead{0.1$\mu$m Sil.}   \\
 	        & \colhead{1.5$\mu$m A.Ol.}  & \colhead{1.5$\mu$m A.Py.}  & \colhead{1.5$\mu$m Fors.}  & \colhead{1.5$\mu$m Enst.}  & \colhead{1.5$\mu$m Sil.}   \\
  	        & \colhead{6.0$\mu$m A.Ol.}  & \colhead{6.0$\mu$m A.Py.}  & \colhead{6.0$\mu$m Fors.}  & ---  & \colhead{6.0$\mu$m Sil.}   \\ } 
\startdata
J0843 & --- & --- & 11.5$^{+0.6}_{-0.6}$ & 5.5$^{+1.2}_{-1.2}$ & 2.8$^{+0.3}_{-0.4}$ \\
 & --- & 68$^{+3}_{-3}$ & 4$^{+2}_{-2}$ & 8$^{+2}_{-2}$ & --- \\
 & --- & --- & --- & --- & --- \\
J0844 & 17$^{+19}_{-13}$ & 60$^{+13}_{-13}$ & $<$3 & $<$8 & --- \\
 & --- & $<$24 & 4$^{+11}_{-4}$ & 15$^{+13}_{-10}$ & $<$5 \\
 & $<$4 & --- & $<$31 & --- & --- \\
RECX-5 & --- & --- & 10.0$^{+0.6}_{-0.5}$ & --- & --- \\
 & 38$^{+9}_{-8}$ & 37$^{+10}_{-10}$ & --- & 12$^{+2}_{-2}$ & 2.0$^{+0.9}_{-0.8}$ \\
 & --- & --- & --- & --- & --- \\
RECX-9 & --- & 20$^{+22}_{-12}$ & 6$^{+1}_{-1}$ & $<$3 & 2.7$^{+0.8}_{-0.9}$ \\
 & --- & 57$^{+14}_{-26}$ & $<$3 & 13$^{+5}_{-5}$ & --- \\
 & $<$7 & --- & $<$9 & --- & --- \\
RECX-11 & --- & --- & --- & --- & --- \\
 & --- & 12$^{+1}_{-1}$ & --- & 3.7$^{+0.4}_{-0.5}$ & --- \\
 & 74$^{+2}_{-3}$ & --- & --- & --- & 7$^{+2}_{-2}$ \\
\\
\\
J0843$^l$ & 4$^{+4}_{-3}$ & --- & 2.7$^{+0.1}_{-0.2}$ & --- & --- \\
 & 89$^{+3}_{-4}$ & --- & --- & 3.0$^{+0.1}_{-0.1}$ & --- \\
 & --- & --- & --- & --- & --- \\
RECX-5$^l$ & --- & --- & 7.6$^{+1.1}_{-1.3}$ & --- & --- \\
 & 52$^{+5}_{-9}$ & 32$^{+5}_{-6}$ & --- & --- & --- \\
 & --- & 6$^{+22}_{-5}$ & --- & --- & $<$3 \\
RECX-11$^l$ & --- & --- & 3.7$^{+0.6}_{-0.5}$ & $<$3 & --- \\
 & 19$^{+8}_{-11}$ & 63$^{+8}_{-8}$ & 3$^{+1}_{-1}$ & --- & 3.3$^{+0.5}_{-0.5}$ \\
 & $<$2 & --- & 4$^{+3}_{-3}$ & --- & $<$6 \\
RECX-9$^l$ & --- & 4$^{+54}_{-4}$ & 3$^{+3}_{-2}$ & $<$6 & 5$^{+3}_{-4}$ \\
 & $<$27 & 70$^{+13}_{-26}$ & 7$^{+6}_{-4}$ & 3$^{+3}_{-2}$ & --- \\
 & $<$28 & 3$^{+34}_{-3}$ & 2$^{+11}_{-2}$ & --- & $<$7 \\
\enddata
\tablecomments{Silicate fit for the objects in $\eta$ Cha using
the Two Layer Temperature Distribution method (J09). 
The values are percentages in mass fraction. If no value is given
for a certain species/size, it means that it is present at 2\% level
or less. The silicate compositions corresponding to the 7-17~$\mu$m range fit
are given in the first part of the table, followed by the 17-35~$\mu$m fit, which are also
marked with $^l$ added to the name. }
\end{deluxetable}

\begin{deluxetable}{lccc}
%\tabletypesize{\scriptsize}
%\rotate
\tablenum{5}
\tablecolumns{4} 
\tablewidth{0pc} 
\tablecaption{Temperatures in the TLTD fit\label{temp-table}} 
\tablehead{
 \colhead{Name}  & \colhead{T$_{rim}$ (K)} & \colhead{T$_{mid}$ (K)} & \colhead{T$_{atm}$ (K)}} 
\startdata
J0843	&	162-1590	&	101-209	&	116-1395	\\
J0844	&	156-1578	&	116-293	&	116-1389	\\
RECX-5	&	159-1564	&	115-178	&	116-362 	\\
RECX-9	&	156-1586	&	114-324	&	115-934 	\\
RECX-11	&	163-1598	&	116-218	&	116-963	 \\
\\
\\
J0843$^l$	&	163-1562	&	59-118	&	60-321	\\
RECX-5$^l$	&	159-1590	&	106-145	&	116-174	\\
RECX-9$^l$	&	159-1595	&	115-189	&	115-221	\\
RECX-11$^l$	&	163-1598	&	116-196	&	115-235	\\
\enddata
\tablecomments{Maximum and minimum temperatures (in K) of the three
components fitted by the TLTD model (inner rim, disk midplane, and disk
atmosphere). See text for details. }
\end{deluxetable}

\clearpage

\begin{deluxetable}{lccc}
%\tabletypesize{\scriptsize}
%\rotate
\tablenum{6}
\tablecolumns{4} 
\tablewidth{0pc} 
\tablecaption{Grain sizes and crystalline fraction (TLTD) \label{sizecryst-table}} 
\tablehead{
 \colhead{Name}  & \colhead{Size (am.) /~$\mu$m} & \colhead{Size (cryst.) /~$\mu$m} & \colhead{Cryst.}} 
\startdata
J0843 & 1.5$^{+0.1}_{-0.1}$ &  0.7$^{+0.1}_{-0.1}$  & 29$^{+2}_{-2}$  \\
J0844 & 0.1$^{+0.3}_{-0.1}$: &  $<$6: &  21$^{+9}_{-9}$:  \\
RECX-5 & 1.5$^{+0.1}_{-0.1}$ &  0.9$^{+0.1}_{-0.1}$  & 22$^{+2}_{-2}$  \\
RECX-9 & 1.2$^{+0.2}_{-0.3}$ &  1.0$^{+0.1}_{-0.2}$  & 19$^{+5}_{-3}$  \\
RECX-11 & 3.4$^{+0.1}_{-0.1}$ &  0.9$^{+0.1}_{-0.1}$  & 6.5$^{+0.4}_{-0.5}$  \\
\\
\\
J0843$^l$ & 1.4$^{+0.1}_{-0.1}$ &  0.8$^{+0.1}_{-0.1}$ &  5.8$^{+0.1}_{-0.2}$ \\
RECX-5$^l$ & 1.8$^{+0.9}_{-0.3}$ &  0.2$^{+0.1}_{-0.1}$  & 9$^{+1}_{-1}$  \\
RECX-9$^l$ & 2$^{+1}_{-1}$: &  1.2$^{+0.6}_{-0.4}$:  & 16$^{+9}_{-7}$:  \\
RECX-11$^l$ & 1.5$^{+0.1}_{-0.1}$ &  2.2$^{+0.6}_{-0.8}$  & 13$^{+3}_{-3}$  \\
\enddata
\tablecomments{Average grain sizes for amorphous and crystalline silicates 
and crystalline fraction (mass fraction of crystalline silicates) from the TLTD fit.
The results corresponding to the 7-17~$\mu$m range fit
are given in the first part of the table, followed by the 17-35~$\mu$m fit, which are also
marked with $^l$ added to the name. The colon (:) indicates uncertain
values in spectra with poor S/N.}
\end{deluxetable}

\clearpage

\end{document}